\documentclass[aps,preprint]{revtex4}
\usepackage{graphicx}
\usepackage{amsmath}

\usepackage{color}

\begin{document}
\title{Potential Energy Landscape of the Apparent First-Order
  Phase Transition between Low-Density and High-Density Amorphous Ice}

\author{Nicolas Giovambattista$^{1,2}$,  Francesco Sciortino$^3$, Francis W. Starr$^4$, Peter H. Poole$^5$}

\affiliation{
 $^1$  Department of Physics, Brooklyn College of the City 
 University of New York, Brooklyn, NY 11210 USA\\
$^2$ Ph.D. Programs in Chemistry and Physics,
     The Graduate Center of the City University of New York, New York, NY 10016\\ 
$^3$ Dipartimento di Fisica and CNR-ISC, Universita di Roma La Sapienza, \\
 Piazzale Aldo Moro 2, I-00185 Rome, Italy     \\
$^4$ Department of Physics, Wesleyan University, Middletown, \\
     Connecticut 06459, United States     \\
$^5$ Department of Physics, St. Francis Xavier University,
     Antigonish, Nova Scotia B2G 2W5, Canada     \\
}

\date{\today}

\begin{abstract}
  The potential energy landscape (PEL) formalism is a valuable approach within statistical
  mechanics for describing supercooled
  liquids and glasses.  Here we use the PEL formalism and
  computer simulations to study the pressure-induced transformations
  between low-density amorphous ice (LDA) and high-density amorphous ice (HDA) at
  different temperatures.  We employ the ST2 water model for which the
  LDA-HDA transformations are remarkably sharp, similar to what is observed in
  experiments, and reminiscent of a first-order phase transition.
Our results are consistent with the view that LDA and HDA configurations 
are associated with two distinct regions (megabasins) of the PEL that are 
separated by a potential energy barrier.  
%We show that the assignment of LDA and HDA to distinct megabasins is plausible regardless of whether the configurations are prepared by temperature quenching the liquid, or compression/decompression of the amorphous solid at fixed temperature.
%%%%%%%%%%%%%%5
  %The LDA and HDA megabasins correspond to two separate {\it families} of
  %configurations.  For example, (i) the starting LDA of the
  %compression/decompression cycle and (ii) the LDA form recovered after
  %decompression of HDA are microscopically different, yet they belong
  %to the same LDA megabasin in the PEL.  
  %% 
At higher temperature, we find that low-density liquid (LDL) configurations are
 located in the same megabasin as LDA, and that high-density liquid (HDL)
 configurations are located in the same megabasin as HDA.  We show that the 
pressure-induced LDL-HDL and LDA-HDA transformations occur along paths that
 interconnect these two megabasins, but that the path followed by the liquid is different than the path followed by the amorphous solid.
    %Amorphous solid
    %configurations (LDA or HDA) are located deep in their respective megabasins 
    %while liquid configurations occur at higher energy on the PEL (REALLY?).
    %It follows that the pressure-induced LDA-HDA and LDL-HDL both correspond 
    %to the system evolving from one megabasin to another but the paths
    %followed in the PEL along the LDL-HDL and LDA-HDA transformations are different.
% Our results are 
%     also consistent with the view that the low-density (LDL) and
%    high-density liquid (HDL) configurations belong to the LDA and HDA
%    megabasins, respectively.  However, at least for the accessible fast
%    compression rates, LDA and LDL (HDA and HDL) sample different
%    regions of the LDL/LDA (HDL/HDA) megabasin. 
%%%%%%
%Although we generally expect
%a close relation between the basins of glassy forms and the liquid,
%  the regions of the LDA/LDL and HDA/HDL megabasins sampled 
%by the glass and the liquid at these high quench
%  rates are not the same as those accessed by the equilibrium LDL and HDL, 
%respectively.
%%%%
  At higher pressure, we also study the liquid-to-ice-VII first-order phase
  transition, and find that the behavior of the PEL properties across this
  transition are qualitatively similar to the changes
  found during the LDA-HDA transformation.  This similarity supports the
  interpretation that the LDA-HDA transformation is a first-order-like phase
  transition between out-of-equilibrium states.
%%%%%
  Finally, we compare the PEL properties explored during the LDA-HDA
  transformations in ST2 water with those reported previously for SPC/E
  water, for which the LDA-HDA transformations are rather smooth.  This comparison illuminates previous work showing that, at accessible computer times scales, a
  liquid-liquid phase transition occurs in the case of ST2
  water, but not for SPC/E water. 
\end{abstract}

\maketitle

\section{Introduction}

Water is one of the most complex liquids, exhibiting many anomalous thermodynamic
and dynamical properties (see e.g.,
Refs.~\cite{pabloReview2003,pabloGene}).  In the glassy state, water is
also a complex
substance~\cite{pabloReview2003,mishimaStanleyNature,angellReview,LGreview}.
Notably, amorphous solid water can be found in at least two different forms, low-density amorphous ice (LDA) and high-density amorphous ice (HDA), having very different properties.  For example, the densities of LDA and
HDA differ by $\approx 20\%$~\cite{mishimaNature1984, mishimaNature1985,
  mishimaVol,katrinRevers2008_1, katrinRevers2008_2}.  A large number of
experiments indicate that LDA and HDA can be interconverted via many
thermodynamic paths, such as isobaric heating and isothermal
compression/decompression processes (see e.g.
Refs.\cite{pabloReview2003,mishimaStanleyNature,angellReview,LGreview}).
The LDA-HDA transformation, between properly annealed LDA and HDA forms,
is rather sharp and reversible, and exhibits many of the characteristics
of a first-order phase transition~\cite{mishimaNature1985,mishimaJPC1994,katrin2006,klotzPRL2005,andersson2002}.  The explanation of this unusual behavior requires
answers to fundamental questions of statistical mechanics, such as how
to properly define or interpret a ``phase-transition'' between
out-of-equilibrium glassy states.

In this work, we address these questions by studying the LDA-HDA
transformations in water, and the relationship of these glasses with the
liquid state, using the potential energy landscape (PEL) approach (see
e.g. Ref.~\cite{stillinger95}).  The PEL is a statistical mechanical
approach that formally separates the configurational contributions to
the partition function into contributions from local energy minima (``inherent structures"), 
and vibrational excitations within the basins of attraction surrounding these minima.
Consequently, the PEL formalism has been used extensively
to study supercooled liquids and
glasses~\cite{pabloStill,francescoReview,harrowellEdiger2012}, and other
equilibrium systems~\cite{daviesBook}, where such a separation between
vibrational and configurational degrees of freedom arises naturally.

Specifically, for a system of $N$ particles, the PEL is the
hyper-surface in $(3N+1)$-dimensional space defined by the potential
energy of the system as function of the $3N$ coordinates,
$U(\vec{r}_1, \vec{r}_2,....\vec{r}_N)$.  At any given time $t$, the
system is represented by a single point in the PEL defined by the
particle coordinates $\{\vec{r}_1(t), \vec{r}_2(t),....\vec{r}_N(t)\}$.
As time evolves, the representative point of the system
moves, sampling different regions of the PEL.  
In the PEL approach, the
thermodynamic~\cite{francescoReview,sastriNature} and
dynamic~\cite{mossa,lanave,scalaNat,francisINM} properties of the
system can be defined in terms of the topography of the PEL regions
being sampled. 
The topography of the PEL can be rather complex with
  comparatively shallow basins residing within deeper and broader
  ``megabasins"~\cite{stillinger95,heuwer1,heuwer2}.  
  In the case of
  supercooled liquids, the free energy can be expressed in terms of
  three basic properties of the PEL~\cite{francescoReview,sastriNature,pabloStill,stillinger95}: 
(i) the average energy $E_{IS}$ of the inherent structures (IS) associated 
with the basins sampled in equilibrium;  
(ii) the number of IS having a given value of $E_{IS}$;  and,
(iii) the average curvature of the basins associated with each IS.

The behavior of glassy and liquid water are necessarily related, 
and several scenarios have been proposed to explain their 
unusual properties within a common framework (see
e.g.~\cite{rev25,PooleNature92,rev38,rev23}).  One of the more widely
accepted explanations is based on the idea that LDA and HDA are the
glass counterparts of two liquids, low-density (LDL) and high-density
liquid (HDL), that are separated by a first-order phase
transition~\cite{PooleNature92,mishimaStanleyNature,poole1993}.  This
liquid-liquid phase transition (LLPT) ends at a liquid-liquid critical point (LLCP)
recently estimated to exist at a temperature $T_{LLCP} \approx 223$~K and
pressure $P_{LLCP} \approx 50$~MPa, based on experiments and
theory~\cite{mishimaVol,FAeos}.  The LLPT hypothesis was originally
proposed on the basis of computer simulations using the ST2 water
model~\cite{PooleNature92}, leaving the validation of the hypothesis for
experiments. Unfortunately, crystallization makes experimental
verification challenging, and thus the hypothesis remains a point of
debate.  However, there is experimental evidence supporting the existence
of a LLPT in water~\cite{winkel2011,thomasReviewTgHDA,mishimaLLCP2000}.
Indeed, most of the evidence supporting the LLPT hypothesis is from
studies involving glassy water.  A LLPT has been
directly observed in experiments on other substances such as
phosphorus~\cite{Katayama} and cerium~\cite{Cadien}, demonstrating the
possibility of such a scenario.  Furthermore, computer simulations of
atomistic (see e.g.~\cite{joel,limei,jagla,francese}), 
nanoparticle~\cite{francisDNA},  and molecular
systems including modified water 
models~\cite{francisNews,smallenburg14,smallenburg15} have shown
the possibility for a LLPT in the equilibrium (as opposed to metastable)
region of the phase diagram, establishing that a LLPT can exist as a
proper phase transition in the thermodynamic limit.

In order to probe the apparent first-order transitions between LDA and
HDA states, we employ the ST2 model. This choice is important since,
contrary to other models such as the SPC/E and mW
models~\cite{davidPNAS2014,MySciRep,yoPhaseDiag}, the ST2 model
reproduces the sharp LDA-HDA transformation observed in experiment.  
The behavior of ST2 water in glassy states has been
recently characterized in detail and it has been shown that it is in
qualitative agreement with experiments~\cite{chiu1,chiu2}.
Specifically, in the case of the ST2 model, the density $\rho$ has been observed to change abruptly with little change in the pressure $P$ [{\it i.e.} $\left( \partial P/ \partial \rho \right)_T \approx 0$] during the LDA-HDA transformation; see e.g. Refs.~\cite{chiu1,MySciRep,jessina}.

In the present work, we explore the LDA-HDA transformation in ST2 water using the PEL formalism in order to clarify the thermodynamic differences between the LDA and HDA forms, and to assess the extent to which it may be appropriate to refer to this transformation as a first-order-like phase transition.  As we show below, our results are consistent with the view that LDA and HDA occupy distinct megabasins of the PEL, and that the transformations 
between LDA and HDA
 exhibit a number of behaviors observed in well-defined equilibrium first-order phase transitions.

The structure of this manuscript is as follows: In Sec.~\ref{methods} we
describe the computer simulation details and methods employed.  In
Sec.~\ref{LDAHDA} we discuss the changes in the PEL properties of ST2
water during the LDA-HDA transformations.  We compare our results for ST2 water 
to those reported previously for SPC/E
water in section~\ref{compare}.  We compare the LDA-HDA transformation and the liquid-to-ice-VII
first-order phase transition using the PEL formalism in
Sec.~\ref{liqIce}.  The regions of the PEL sampled by the liquids and
glasses are compared in Sec.~\ref{liqglassPEL}. A summary and
discussion, including a simple model of the PEL for polyamorphic water,
is presented in Sec.~\ref{summary}.  In the Appendix, we study the
effects of reducing the compression/decompression rates on our results.
Additional material is included as  suplementary information (SI) where 
results at different temperatures are compared.

\section{Simulation and Analysis Methods}
\label{methods}

We perform extensive out-of-equilibrium molecular dynamics (MD)
simulations of water using the ST2 model~\cite{ST2model}.  Long-range
(electrostatic) interactions are treated using the reaction field
technique~\cite{reacField}.  The details of the MD procedure and
our implementation of the ST2 model are identical to that described 
in Ref.~\cite{poole2005}.  Ref.~\cite{chiu1} contains a
complete analysis of the thermodynamic and structural changes
accompanying the pressure-induced LDA-HDA transformations to be
analyzed in this work using the PEL formalism.  

To summarize the
thermodynamic procedure briefly, an LDA configuration is prepared by cooling
equilibrium liquid water at $P=0.1$~MPa from $T_0=350$~K down to the
chosen temperature $T$ for compression/decompression, using a cooling rate of
$q_c=30$~K/ns.  This preparation method corresponds to the experimental process used to produce the LDA form known as hyperquenched glassy water (HGW),
although we use a faster cooling rate than experiments;  see discussions
in Refs.~\cite{chiu1,chiu2,jessina}.  Our HGW configuration is then
compressed isothermally, producing a sample of HDA.  The resulting HDA form is then
decompressed (at the same temperature $T$) leading to a new LDA sample.
When subjected to sufficiently negative pressure, this LDA sample fractures.
In order to compare our HGW sample and the LDA form we
obtain by decompression of HDA, we also subject the HGW configuration to increasingly negative pressure, starting from $P=0.1$~MPa, until it also fractures.  

Our
compression/decompression rate is $q_P=300$~MPa/ns, which leads to sharp
LDA-HDA transformations, as observed experimentally using much slower
rates~\cite{chiu1,MySciRep,jessina}.  For each compression/decompression
temperature, we perform $10$ runs starting from independently generated starting
configurations to account for sample-to-sample variations in the
non-equilibrium state. In all cases, we use a cubic box with $N = 1728$
water molecules with periodic boundary conditions.

During the compression/decompression runs, configurations are saved
every $10$~MPa and the corresponding IS
%minimum energy structures, commonly called inherent structures (IS),  
are obtained using the conjugate gradient algorithm~\cite{conjGrad}.
The virial expression for the pressure 
 at the IS configuration defines the IS pressure,
$P_{IS}$ (see e.g., Ref.~\cite{nymand}). %% eqns 18 and 37
 The curvature of a basin near the IS minimum is
quantified by the shape function ${\cal S}_{IS}$.  
${\cal S}_{IS}$ is obtained from the set of eigenvalues 
$\{\omega_i^2\}$ (where $i=1$ to $6N$)
of the Hessian matrix evaluated at the IS configuration (see e.g.
Ref.~\cite{francescoReview} and references therein):
\begin{eqnarray}
{\cal S}_{IS} =  \frac{1}{N} \sum_{i=1}^{6N-3} \ln \left( \frac{\hbar \omega_i}{A_0}   \right).
\end{eqnarray}
Here $\omega_i$ is the frequency of vibrational mode $i$, and 
$\hbar= h/2 \pi$ where $h$ is Planck's constant.  The constant $A_0=1$~kJ/mol is included so that the argument of the $\ln$ function is dimensionless. 

The same definition for ${\cal S}_{IS}$ was employed in
Ref.~\cite{yoSPCE-LDAHDA} for the case of SPC/E water, allowing for a
direct comparison of the present results with those reported in Ref.~\cite{yoSPCE-LDAHDA}.
While $P_{IS}$ and $E_{IS}$ are calculated for $10$ independent runs,
the shape function ${\cal S}_{IS}$ is calculated for only $2$ of these
runs, owing to the computational expense of evaluating the eigenvalues $\{\omega^2_i\}$.  Technical details on the evaluation of the
Hessian can be found in
Ref.~\cite{matharooPoole,sastryFrancesco94} where the density of states
of water in equilibrium is reported.

 The PEL properties $E_{IS}$, $P_{IS}$, and ${\cal S}_{IS}$ 
are fundamental quantities in the PEL formalism~\cite{francescoReview}. 
For example, for a low-temperature liquid in equilibrium (or metastable
equilibrium), knowledge of $E_{IS}$ and ${\cal S}_{IS}$ as a function of $V$ and $T$ is sufficient to quantify key contributions to the free energy.  In this case, the Helmholtz free energy $F$ can be written as,
\begin{eqnarray}
F=E_{IS}-T S_{conf} + E_{vib} - T S_{vib},
\label{FFF}
\end{eqnarray}
where $S_{conf}$ and $S_{vib}$ are the configurational and vibrational entropies and 
$E_{vib}$ is the average energy of the system due to vibrational motion in the PEL
 around the IS.  In the harmonic approximation,
\begin{eqnarray}
S_{vib} = f(T,N) - k_B {\cal S}_{IS},
\label{SIS}
\end{eqnarray}
where $f(N,T)$ is a function only of temperature
and number of molecules~\cite{mossaPELmolec}, and
\begin{eqnarray}
E_{vib} = 3 N k_B T
\label{Evib}
\end{eqnarray}
In the case of water, there can also be significant anharmonic contributions 
\cite{waterAnharm}.
An extension of the PEL formalism to quasi-equilibrium liquids shows
that $E_{IS}$, $P_{IS}$, and ${\cal S}_{IS}$ 
are sufficient to write an analogous
expression for the free energy of out-of-equilibrium states, from which
the thermodynamic properties of the liquid can be
predicted~\cite{mossaOOE_1,mossaOOE_2,tartagliaOOE}.  In this case,
however, knowledge of the PEL properties sampled in
equilibrium is necessary~\cite{francescoReview}.

\section{The LDA-HDA Transformation in the PEL Formalism}
\label{LDAHDA}

We focus on the three fundamental properties of the PEL ($P_{IS}$,
$E_{IS}$, and ${\cal S}_{IS}$) sampled by the system during the
compression-induced LDA-to-HDA transformation and the decompression-induced
HDA-to-LDA transformation.  Doing so allows us to characterize the PEL basins associated with these states, and to determine the overall topography of the PEL for this system.

%\subsection{LDA-HDA transformation at low temperatures}
%\label{compDecomp}

\begin{figure}[ht]	% Fig1
\centerline  {
\includegraphics[width=7.0cm]{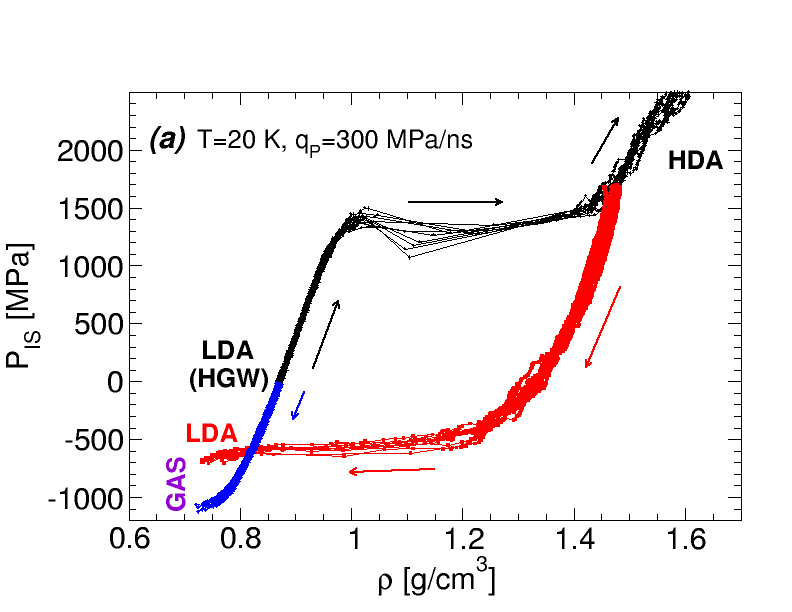}
\includegraphics[width=7.0cm]{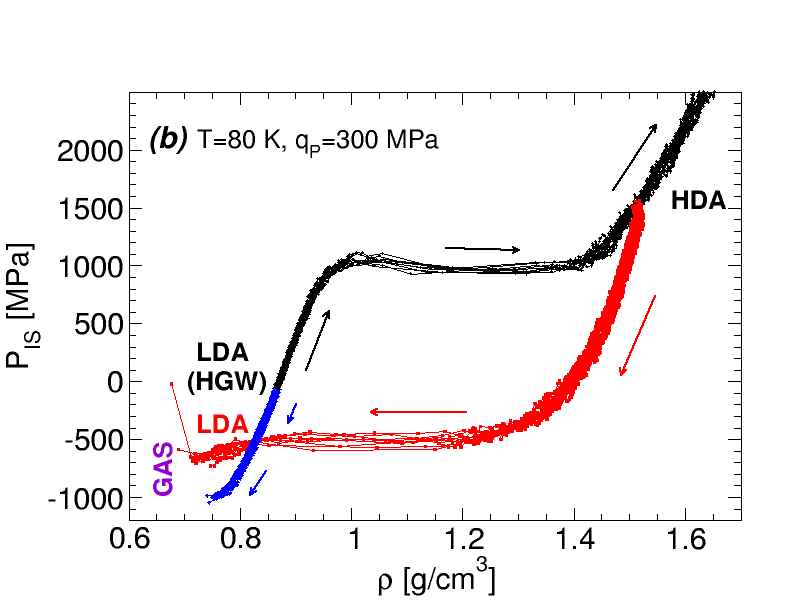}    
}
\centerline  {
\includegraphics[width=7.0cm]{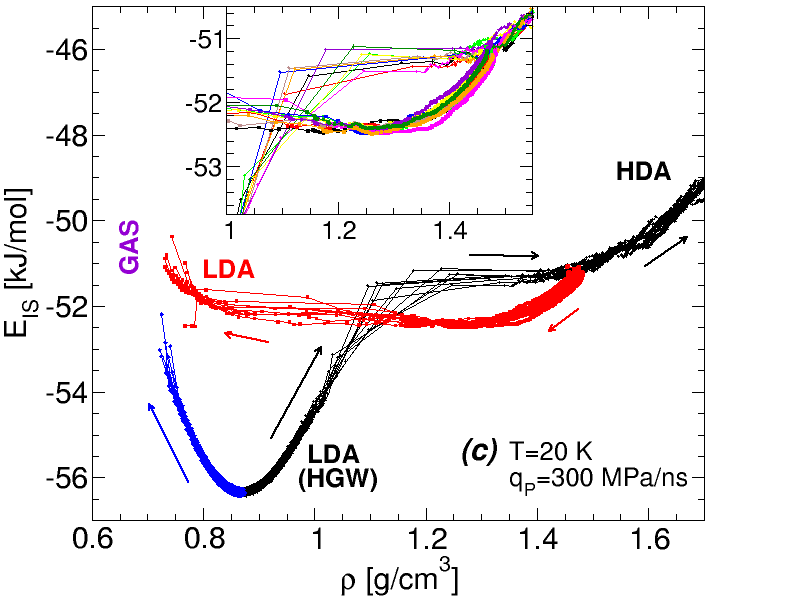}
\includegraphics[width=7.0cm]{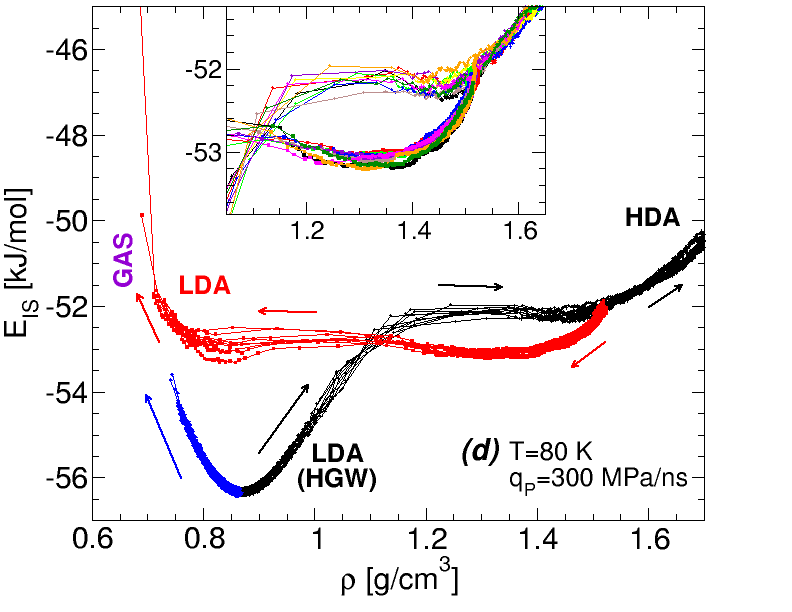} 
}
\centerline  {
\includegraphics[width=7.0cm]{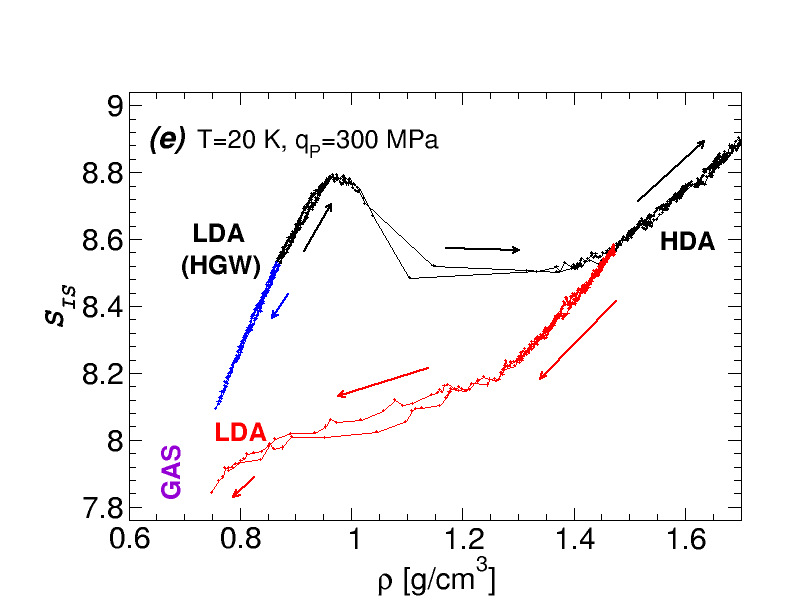}
\includegraphics[width=7.0cm]{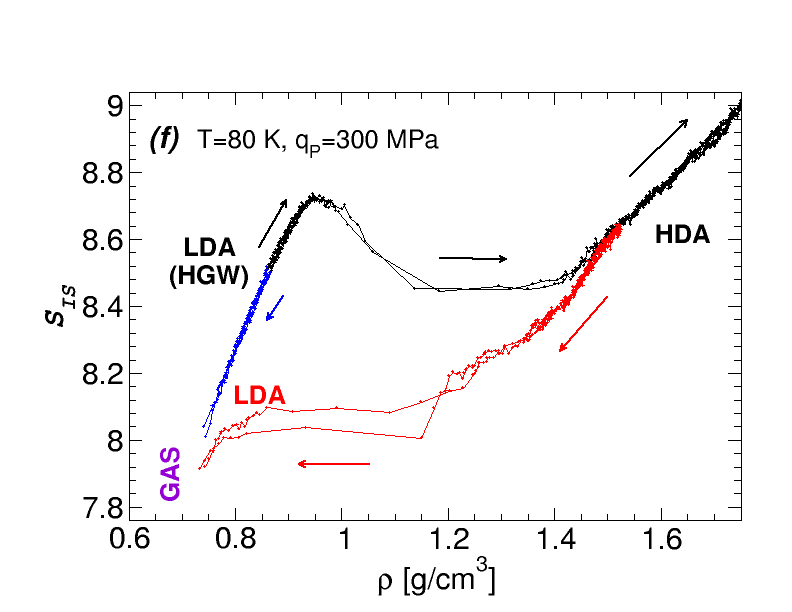} 
}
\caption{(a)(b) Pressure $P_{IS}$, (c)(d) energy $E_{IS}$, and (e)(f) 
shape function ${\cal S}_{IS}$ of the inherent structures
 sampled during the pressure-induced LDA-HDA transformations at $T=20$~K (left column)
 and $80$~K (right column).  Black and red lines correspond to compression [LDA(HGW)-to-HDA]
 and decompression [HDA-to-LDA-to-gas] runs, respectively; blue lines 
correspond to the decompression of HGW (generated by cooling the liquid at $P=0.1$~MPa).
At these temperatures, ST2 glassy water does not crystallize ($P<6000$~MPa).  
Results for $P_{IS}$ and $E_{IS}$ are from $10$ independent compression/decompression runs; 
$2$ independent runs are used for the calculation of ${\cal S}_{IS}$.
%NOT OBSERVED IN SPCE:  (1) nice two minima in $Eis$. (2) nice van 
%der Walls-like loops in $P_{IS}$ and ${\cal S}_{IS}$.
} 
\label{PEL-LDAHDA}
\end{figure}

First we examine the LDA-to-HDA transformation at very low temperatures.
%in which no crystalline ice is found during the compression runs for $P\leq 6000$~MPa~\cite{chiu1}. 
 Figs.~\ref{PEL-LDAHDA}(a)-(f) show
$P_{IS}$, $E_{IS}$, and ${\cal S}_{IS}$ as a function of $\rho$ at two
representative temperatures in the glass state, $T=20$ and
  $T=80$~K, during the LDA-to-HDA transformation.
  The behavior of $P_{IS}(\rho)$,
$E_{IS}(\rho)$, and ${\cal S}_{IS}(\rho)$ does not differ significantly
between these two $T$.  Hence we will focus our discussion on the case
$T=80$~K.

In Fig.~2(b) of Ref.~\cite{chiu1}, the behavior of the pressure $P(\rho)$ during the LDA-to-HDA
transformation at $T=80$~K is shown
for the same $10$ compression runs for which $P_{IS}(\rho)$ is shown here in
Figs.~\ref{PEL-LDAHDA}(b).  The LDA-to-HDA transformation during
compression at $80$~K occurs at a pressure $P_{\rm LDA-to-HDA}$ which ranges between $1050$ and $1200$~MPa.  Within the
LDA regime ($P<P_{\rm LDA-to-HDA}$) and the HDA regime ($P>P_{\rm LDA-to-HDA}$), we note that $P$ varies rapidly with $\rho$ and is approximately linear, indicating that in these regimes the system responds to volume changes like a stiff elastic solid.
%%Indeed, since the change in density within the LDA and HDA states in this pressure range are small, $\Delta \rho \approx 0.1-0.15$~g/cm$^3$, one finds that $P \propto \rho$ as well. 
%This behavior indicates that within
%the LDA and HDA regimes, compression leads to approximately elastic
%deformations. 
% {\color{red} The previous sentence is unclear to me. How
%  can something be linearly dependent both in x and in 1/x (V and rho)
%  ?} 
These elastic regimes correspond to approximately $\rho < 0.95$ for LDA and $\rho>1.45$~g/cm$^3$ for HDA.
As shown in Fig.~\ref{PEL-LDAHDA}(b), we find that the behavior of $P_{IS}(\rho)$ in these density ranges is similar to that observed for $P(\rho)$ in Ref.~\cite{chiu1}.

A significant difference between $P(\rho)$ and $P_{IS}(\rho)$ is that $P$ is a monotonic function of $\rho$, while $P_{IS}$ is not.  That is, during the LDA-to-HDA transformation
 (occurring in the range $0.95 < \rho < 1.45$~g/cm$^3$), $P_{IS}(\rho)$ exhibits a van der Waals-like loop reminiscent of a first-order phase transition;  see Fig.~\ref{PEL-LDAHDA}(b).
This van der Waals-like loop in $P_{IS}(\rho)$ becomes
  more evident at $T=160$~K [see Fig.~\ref{PEL-LDAHDAICE}(a)], {\it i.e}. as
  the liquid phase is approached from the glass state, and it vanishes
  at $T>T_{LLCP}$ [see Fig.~\ref{PEL-LDAHDAICE}(b)].   
Since our system is glassy and therefore not in equilibrium, we cannot say that 
a van der Waals loop in
$P_{IS}(\rho)$ indicates a thermodynamic instability.  At the same time, it is 
interesting that 
during the LDA-to-HDA transformation we find
$\left( \partial P_{IS}/ \partial \rho \right)_T < 0$, in analogy with
the condition of instability for an equilibrium system,
$\left( \partial P/ \partial \rho \right)_T < 0$.  We
note that for $T \rightarrow 0$~K, these two inequalities become
identical, since in this limit
$P_{IS} \rightarrow P$ because the vibrational
contributions to the pressure vanish~\cite{francescoReview}.

\begin{figure}[ht]      % Fig2
\centerline  {
\includegraphics[width=7.0cm]{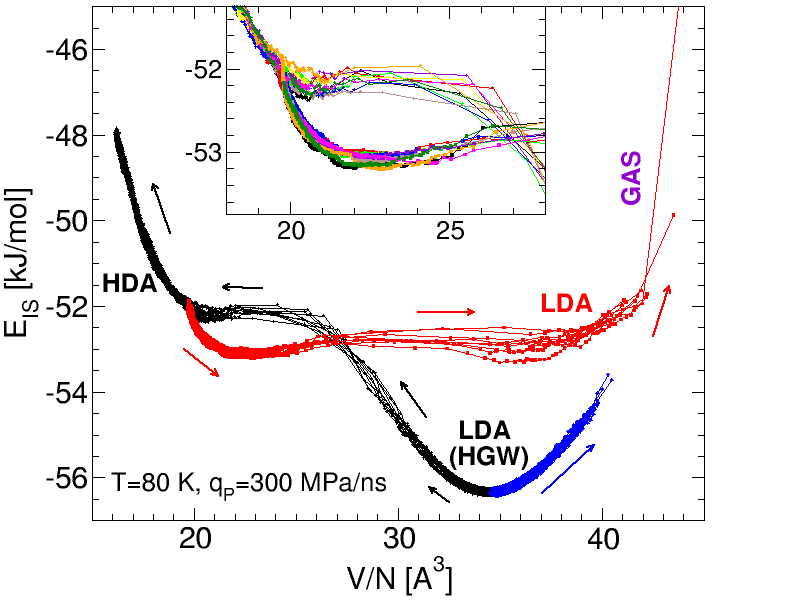}
}
\caption{Volume-dependence of the IS energy during the pressure-induced LDA-HDA 
transformations at $T=80$~K [data is taken from Fig.~\ref{PEL-LDAHDA}(c)]
Black and red lines correspond to compression [LDA(HGW)-to-HDA]
 and decompression [HDA-to-LDA-to-gas] runs, respectively; blue lines
correspond to the decompression of HGW (generated by cooling the liquid at $P=0.1$~MPa).
}
\label{Eis-v}
\end{figure}

\begin{figure}[ht]      % Fig3
\centerline  {
\includegraphics[width=7.0cm]{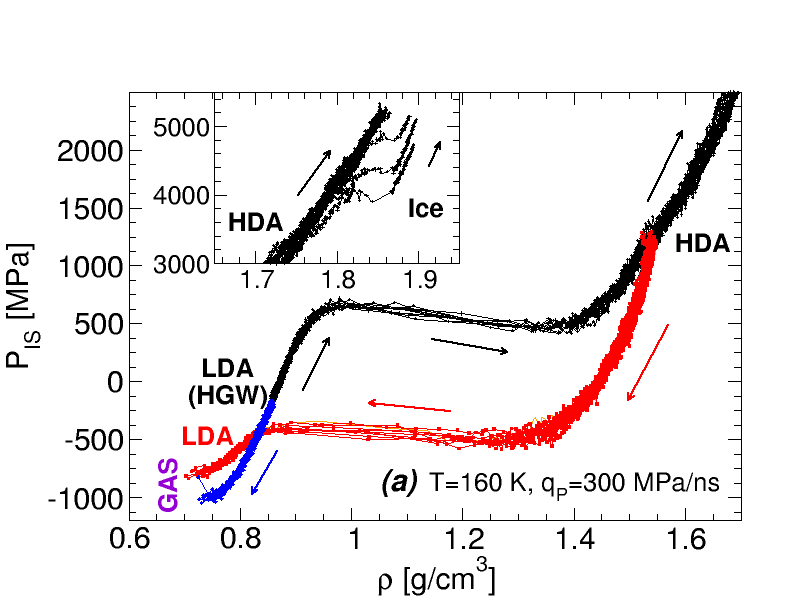}
\includegraphics[width=7.0cm]{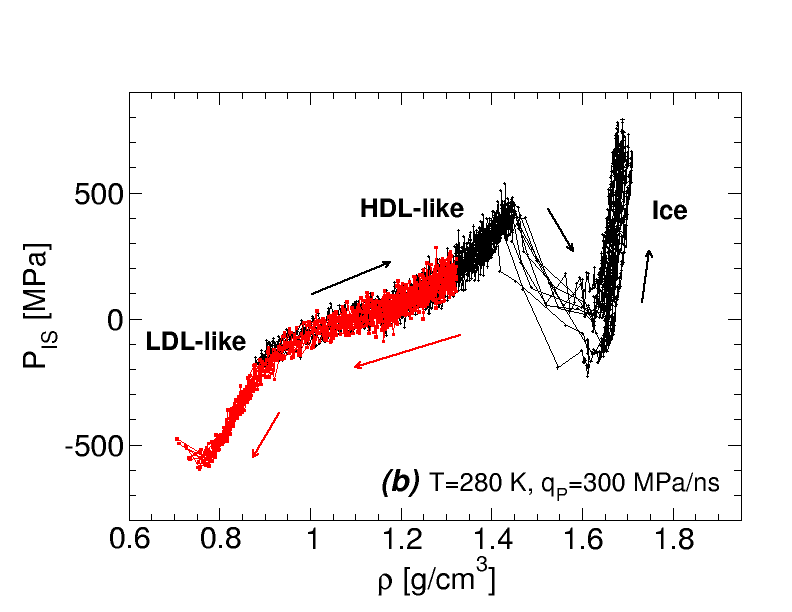}
}
\centerline  {
\includegraphics[width=7.0cm]{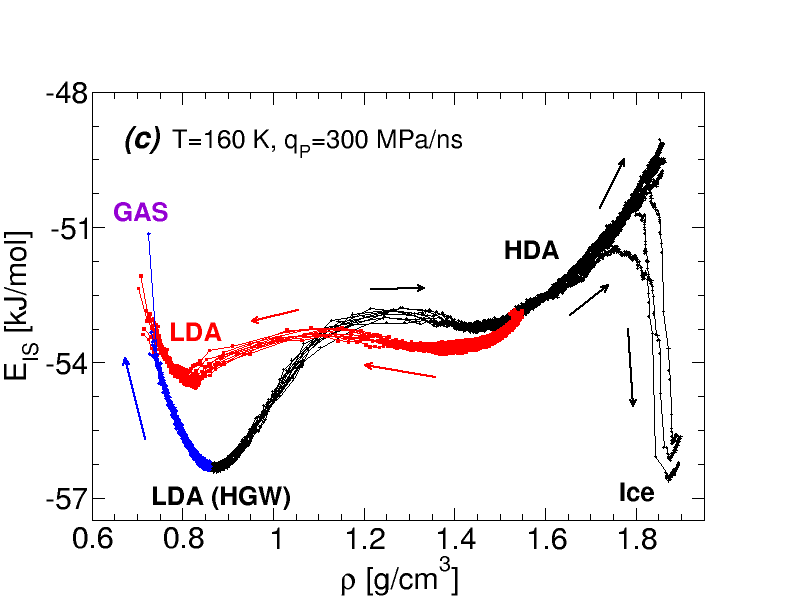}
\includegraphics[width=7.0cm]{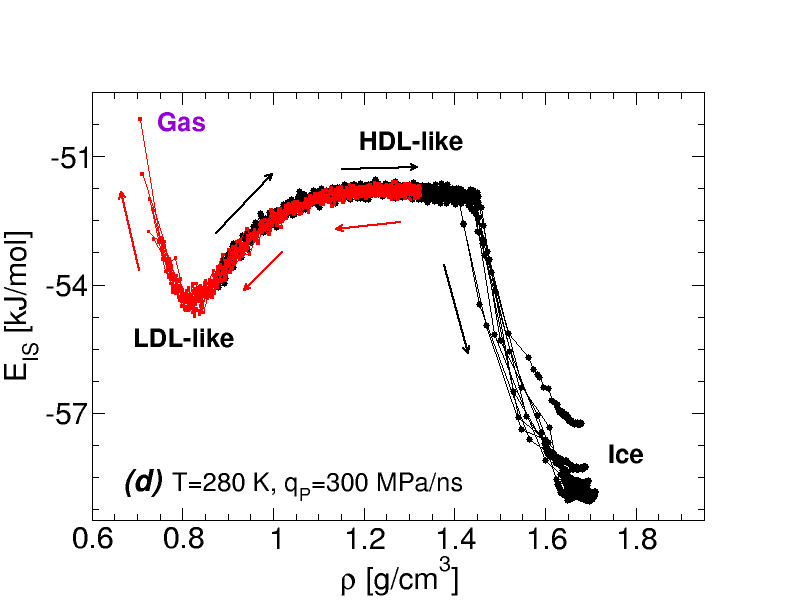}
}
\centerline  {
\includegraphics[width=7.0cm]{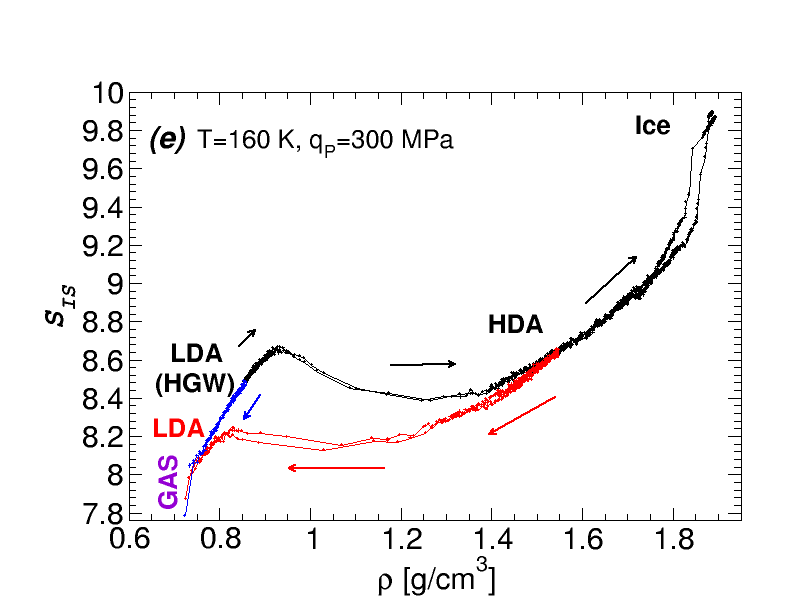}
\includegraphics[width=7.0cm]{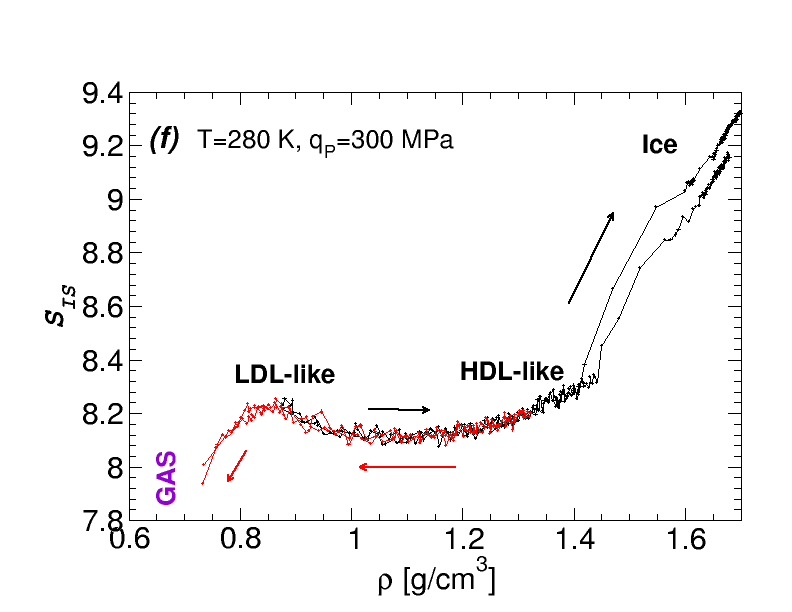}
}
\caption{(a)(b) Pressure $P_{IS}$, (c)(d) energy $E_{IS}$, and (e)(f) shape 
function ${\cal S}_{IS}$ of the inherent  structures 
sampled during the pressure-induced LDA-HDA transformations at $T=160$~K (left column) and 
(LDL-like)-(HDL-like)  transformations at $280$~K (right column) 
[note that $T_{LLCP} \approx 245$~K]. At these temperatures, 
 ST2 water crystallizes at high densities. 
Black and red lines correspond to compression [LDA(HGW)-to-HDA-to-Ice ($T=160$~K) and 
(LDL-like)-to-(HDL-like)-to-Ice ($T=280$~K)] and decompression 
[HDA-to-LDA-to-gas ($T=160$~K) and (HDL-like)-to-(LDL-like)-to-gas ($T=280$~K)] runs, 
respectively; blue lines at $T=160$~K
correspond to the decompression of HGW (generated by cooling the liquid at $P=0.1$~MPa).
Results for $P_{IS}$ and $E_{IS}$ are from $10$ independent compression/decompression runs;
$2$ independent runs are used for the calculation of ${\cal S}_{IS}$. 
%NOT OBSERVED IN SPCE:  (1) at $T=160$~K, nice two minima in $Eis$. (2)  at $T=280$~K, no minimum in $E_{IS}$ associated to HDA; but nice minima in $E_{IS}$ for ice. (3) nice van der Walls-like loops in $P_{IS}$ FOR BOTH HDL-ice ($T=280$~K) AND LDA-HDA ($T=160$~K).
}
\label{PEL-LDAHDAICE}
\end{figure}

We next examine the behavior of $E_{IS}$ during the LDA-to-HDA transformation
at $T=80$~K, shown in Fig.~\ref{PEL-LDAHDA}(d).  
In the following analysis we assume that $E_{IS}$ contributes to $F$ as 
described by Eq.~\ref{FFF}.
 In this case, however, $S_{conf}$
 is considered to be an out-of-equilibrium configurational entropy
 that depends on the glass preparation.  This implies that $F$ is assumed 
to be an additive function of   $E_{IS}$ and $-k_B T {\cal S}_{IS}$.       
 We also note that the inequality 
$\left( \partial^2 F/\partial V^2 \right)_{T}<0$ identifies conditions at which a 
system is thermodynamically unstable as a single homogeneous phase.  
As shown in Fig.~\ref{Eis-v},  the curvature of $E_{IS}$ as a function of $V$ 
is positive in the density ranges associated with the elastic regimes of both LDA and HDA, 
consistent with the mechanical stability of the system in these two forms.  
However, during the transformation of LDA 
to HDA (in the range $0.95 < \rho < 1.45$~g/cm$^3$), $E_{IS}$ exhibits a pronounced negative 
curvature, {\it i.e.} $\left( \partial^2 E_{IS}/\partial V^2 \right)_{T}<0$.  Our data for $E_{IS}$
 therefore indicate that the influence of the PEL on the thermodynamic 
behavior of the system is to introduce a region of instability between the LDA and HDA regimes.

We also note that $E_{IS}$ exhibits a weak maximum during the LDA-HDA transformation. 
Hence, the present results are consistent with the PEL of ST2 water containing
two megabasins, one for LDA and another for HDA, separated by {\it potential energy}
 barriers. We stress that such barriers separating the LDA and HDA megabasins
 are not necessary for the system to become unstable. 
In other words, as argued in the previous paragraph (based on Eq.~\ref{FFF}), 
 the negative {\it curvature} of $E_{IS}$ (which does not need to be accompanied by 
{\it potential energy} barriers) may be sufficient to  
introduce an instability [$\left( \partial^2 F/\partial V^2 \right)_{T}<0$] 
between the LDA and HDA states.
  
The behavior of ${\cal S}_{IS}$ during the LDA-to-HDA transformation 
at $T=80$~K is shown in Fig.~\ref{PEL-LDAHDA}(f).  
During the elastic compression of LDA and HDA (for $\rho < 0.95$ and $\rho>1.45$~g/cm$^3$), ${\cal S}_{IS}$ increases monotonically.
In other words, in regimes of elastic deformation, the system explores basins in the PEL that become narrower as density increases.
However, the overall behavior of ${\cal S}_{IS}$ as a function of $V$ across the LDA-to-HDA transformation is non-monotonic.
At the beginning of the LDA-to-HDA transformation, ${\cal S}_{IS}$ decreases abruptly, 
indicating that ``wider'' basins (relative to LDA) become available to the system in the transformation zone.
We also note that 
$\left (\partial^2 {\cal S}_{IS}/ \partial V^2 \right )_T > 0$ throughout much of the LDA-to-HDA transformation region.  
Within the harmonic approximation of the PEL formalism (see Eqs.~\ref{FFF} and \ref{SIS}), the positive curvature of ${\cal S}_{IS}$ will act to reduce the instability introduced by the negative curvature of $E_{IS}$ noted above.  However, in terms of their respective contributions to $F$ according to Eq.~\ref{FFF}, the relative variation of $E_{IS}$ over the LDA-to-HDA transformation range is approximately 20 times larger than for $k_BT{\cal S}_{IS}$ when $T=80$~K.  Hence the curvature of $E_{IS}$ dominates over that contributed by ${\cal S}_{IS}$ in determining the overall curvature of $F$.

To summarize, Figs.~\ref{PEL-LDAHDA}(a)-(f) show three characteristic features of the LDA-to-HDA transformation in the PEL:  (i) a van der Waals-like loop in $P_{IS}$; (ii) negative curvature in $E_{IS}$; and (iii) non-monotonic variation of ${\cal S}_{IS}$.  
These features are all consistent with behavior analogous to a first-order 
phase transition in an equilibrium system,
as we show in Sec.~\ref{liqIce}. 
We note that these characteristic features, found here 
at $T \leq 80$~K, become more pronounced as
$T$ increases within the range of $T$ in which the system remains glassy.
For example, see the results for
$T=160$~K in the left column of Fig.~\ref{PEL-LDAHDAICE} for
$\rho<1.6$~g/cm$^3$. See also the SI where we include $E_{IS}(\rho)$, 
$P_{IS}(\rho)$, and ${\cal S}_{IS}(\rho)$ for all temperatures studied.

%%%%% DECOMP
During decompression, the behavior of $P_{IS}(\rho)$, $E_{IS}(\rho)$ and
${\cal S}_{IS}(\rho)$ are consistent with the system moving from the HDA
megabasin back to the LDA megabasin.  For example, at $T=160$~K [Fig.~\ref{PEL-LDAHDAICE}(a), (c), (e)], all three of the characteristic features noted above [points (i), (ii), (iii)] 
are observed during the HDA-to-LDA transformation.
However, these features are less evident, or absent, during the HDA-to-LDA transformations at
$T \leq 80$~K.  As shown in Fig.~\ref{PEL-LDAHDA}, the negative curvature of $E_{IS}$ is less prominent during decompression, and non-monotonic behavior in $P_{IS}$ and ${\cal S}_{IS}$ is barely observed.
This demonstrates that the IS basins of the PEL visited by the system 
during the LDA-to-HDA and HDA-to-LDA transformations are quite different, that is, 
the system follows different trajectories on the PEL.  

%%%%%% HGW vs LDA_recovered
A natural question follows: Is the LDA-to-HDA transformation reversible?  
%One of the main results of Figs.~\ref{PEL-LDAHDA} and
%~\ref{PEL-LDAHDAICE} is related to the reversibility on the LDA-to-HDA
%transformation.  Specifically, the
As shown in Fig.~\ref{PEL-LDAHDA}, the values of $P_{IS}$, $E_{IS}$, and ${\cal S}_{IS}$
at $\rho \approx 0.87$~g/cm$^3$ for (a) the LDA (HGW) used to 
initiate the compression runs (black
lines) and (b) the LDA recovered from HDA via decompression (red lines), are different.
While these two LDA forms are not identical, as shown in
Ref.~\cite{chiu1}, both are structurally similar based
on the corresponding OO, OH, and HH radial distribution functions.
Ref.~\cite{chiu1} argued that both of these LDA forms should be considered to be
different members of the LDA ``family".  The
present results based on the PEL approach support this view. 
A region of negative curvature in $E_{IS}$ (accompanied by a weak maximum) 
is encountered both during compression and decompression, consistent with the view that these trajectories take the system first from the LDA to the HDA megabasin, and then back again to the LDA megabasin.
However, $E_{IS}$ for the recovered
LDA is much higher than for the starting LDA (HGW); see
Figs.~\ref{PEL-LDAHDA}(c), \ref{PEL-LDAHDA}(d), and
\ref{PEL-LDAHDAICE}(c) for $\rho=0.87$~g/cm$^3$.  
This difference suggests that the
recovered LDA form is a highly stressed glass residing higher in the LDA
megabasin than HGW. This would not be surprising given the relatively fast compression
rates that are accessible in
simulations~\cite{davidPNAS2014,chiu1,MySciRep,jessina,martonak}.
To confirm this interpretation, we subjected both the HGW and the recovered LDA forms to large negative pressures, to test if their properties become more similar when they are both brought close to their limits of mechanical tensile stress.  As shown by the blue lines in 
Figs.~\ref{PEL-LDAHDA} and~\ref{PEL-LDAHDAICE}, the PEL properties of HGW do approach more closely those of the recovered LDA at negative pressure.
Also consistent with this view, we show in the
  Appendix that $E_{IS}(\rho)$ for HGW and recovered LDA
  become closer to one another when the compression/decompression rate
  is reduced.

%\subsection{LDA-HDA Transformation in the Presence or Absence of a LLPT}

\section{Comparison of the LDA-HDA Transformation in ST2 and SPC/E water}
\label{compare}

%%%%%%  SCP/E vs ST2
%One may wonder if the results reported for ST2 water are found in other water models.  
In a previous work, a similar PEL study of
the LDA-HDA transformation in water was carried out using the SPC/E
model~\cite{yoSPCE-LDAHDA}.  The computer simulation protocol followed
in Ref.~\cite{yoSPCE-LDAHDA} is similar to the procedure used in the
present work.  In particular, both studies employ the same
compression/decompression and cooling rates, which is crucial for a
proper comparison.  The main difference between the SPC/E and ST2 models is
that a LLPT is accessible in (metastable) equilibrium simulations of ST2 
water~\cite{PooleNature92,pabloLLCPnat2014,yangLLCP,pooleLLCP2013,smallenburg15}
 while it is not in SPC/E water via
ordinary molecular dynamics simulations.  Earlier
studies~\cite{scalaSPCE,sciortinoLLCPspce} suggested that a LLPT may  be
present in SPC/E water at very low $T$, but it is not accessible at the
available computational time scales.  In short,
Ref.~\cite{yoSPCE-LDAHDA} describes the PEL properties explored during
the LDA-HDA transformation if the LLPT is not
accessible to simulations.  The present study corresponds to the case when
the LLPT is accessible.

The main difference between the LDA-HDA transformation results for SPC/E and ST2
water is that in the case of SPC/E water, the evidence (at e.g.
$T\approx 80$~K; see Fig.~2 in Ref.~\cite{yoSPCE-LDAHDA}) for first-order 
phase transition-like behavior in the PEL is significantly weaker.
Specifically, during the LDA-HDA transformations in SPC/E water: 
(i) $P_{IS}$ does not exhibit van der Waals-like loops;
(ii) $E_{IS}$ exhibits only very weak, barely noticeable, negative curvature 
upon compression, and not at all during decompression; and
(iii) non-monotonic behavior in ${\cal S}_{IS}$ is also weak during compression, and absent during decompression.
This pattern of behavior is consistent with 
the absence of an accessible LLPT in
SPC/E water.

We also note that at the compression rates studied, the LDA-HDA
transformation in SPC/E water exhibits a more gradual change of density on
compression/decompression, compared to ST2 water or experimental 
water~\cite{jessina,chiu1,MySciRep}.  This
suggests that the glass phenomenology observed in SPC/E water can be thought of as a ``supercritical" LDA-HDA transformation,
analogous to a liquid-gas transformation at $T>T_c$.  
In contrast, for ST2 water and perhaps for real water, the glass phenomenology 
corresponds to a ``subcritical'' first-order phase transition between LDA and HDA.

A theory for amorphous ices has been presented in
Ref.~\cite{davidPNAS2014} based on theory/computer simulations using the
mW model, a coarse-grain model for water~\cite{mWmodel}.
In the suplementary information of Ref.~\cite{davidPNAS2014}  it is shown that
for a compression rate $q_P=500$~MPa/ns, similar to the compression rate
employed here and in Ref.~\cite{yoSPCE-LDAHDA}, the 
slope of $\rho(P)$ in the LDA state and during the 
LDA-HDA transformation are similar, {\it i.e.}, there is not a sudden change in density 
associated with the LDA-to-HDA transformation. The authors conclude that 
 there is a state at which LDA, HDA, and the liquid coexist at an out-of-equilibrium
triple point. Since the mW model seems to lack the 
sharp change in density associated with the LDA-to-HDA transformation,
 its behavior is closer to the case of SPC/E than ST2 water.
It is thus suggestive that the theory proposed in Ref.~\cite{davidPNAS2014} may
apply only for the case of ``supercritical" LDA-HDA
transformations, such as that which occurs in SPC/E water.

We finally note some of the characteristic features indicative of a
LDA-HDA ``first-order phase transition", found in ST2 water at
$T \leq 160$~K, vanish with increasing temperature as
$T\rightarrow T_{LLCP}$, where $T_{LLCP} \approx 245$~K is the LLCP
temperature of ST2 water~\cite{poolePRL2011}.  For example, at temperatures slightly
above $T_{LLCP}$, we find that the curvature of $E_{IS}(\rho)$ increases 
(becomes less negative) and $P_{IS}(\rho)$ shows no van der Waals-like loop (see SI).
  In particular, the behavior of  $E_{IS}(\rho)$ and  $P_{IS}(\rho)$ 
 becomes closer to that reported in Fig.~2 of
Ref.~\cite{yoSPCE-LDAHDA} for the case of SPC/E water.  A similar
phenomenology is found at high temperature, in the liquid state.  For
example, the right column of Figs.~\ref{PEL-LDAHDAICE} shows
$E_{IS}(\rho)$, $P_{IS}(\rho)$, and ${\cal S}_{IS}(\rho)$ for the case
of $T=280$~K~$>T_{LLCP}$. At these temperatures, $E_{IS}(\rho)$ and
$P_{IS}(\rho)$ increase smoothly and monotonically with increasing
density (at $\rho<1.45$~g/cm$^3$, where crystallization does not
interfere).

\section{The LDA-HDA transformation Contrasted with the Liquid-to-Ice Transition}
\label{liqIce}

In this section, we study the PEL behavior during the pressure-induced HDA-to-ice (at low
temperature) and liquid-to-ice transformation (at high temperature).  
Refs.~\cite{chiu1,MySciRep,yangLLCP} showed
that ST2 water crystallizes spontaneously at high-pressure during
isothermal compression, or isothermal heating of HDA at high pressure.
The resulting crystal has the structure of ice VII, characterized by
interpenetrating tetrahedral networks.  
The occurrence of spontaneous crystallization in our system provides the opportunity to identify the PEL behavior that is characteristic of an unambiguous first-order phase transition.
We can thus compare 
the behavior of the PEL properties sampled during the
liquid-to-ice first-order phase transition and with that occurring during the LDA-HDA
transformation. 

We first focus on the case of isothermal compression at $T=160$~K; see the left
column of Fig.~\ref{PEL-LDAHDAICE}.  At this temperature, we observe the successive 
LDA-to-HDA and HDA-to-ice transformations upon compression.  The similarities in the behavior of $E_{IS}(\rho)$, $P_{IS}(\rho)$,
and ${\cal S}_{IS}(\rho)$ during these two transformations is striking.  In both the LDA-HDA and HDA-ice transformations, 
(i) $P_{IS}(\rho)$ exhibits a van der Waals loop;
(ii) $E_{IS}(\rho)$ exhibits negative curvature; 
and (iii) a sharp change in ${\cal S}_{IS}(\rho)$ occurs as the transformation proceeds.  
In addition,  the changes observed in the PEL properties during both transformations
are consistent with the system evolving from one megabasin of the PEL to another. Specifically, at
$T=160$~K, a maximum in $E_{IS}$ separates LDA, HDA, and ice indicating that
at this temperature, there are well-separated  LDA, HDA, and ice megabasins 
in the PEL.  
 The
only minor difference is that during the HDA-ice transformation,
${\cal S}_{IS}(\rho)$ increases monotonically (see
Fig.~\ref{PEL-LDAHDAICE}(e) for $\rho > 1.5$~g/cm$^3$), while for the
LDA-HDA transformation ${\cal S}_{IS}(\rho)$ increases
non-monotonically (see Fig.~\ref{PEL-LDAHDAICE}(e) for
$0.85 < \rho <1.7$~g/cm$^3$).  Yet, in both cases, the basins become
narrower [{\it i.e.}, they have a larger ${\cal S}_{IS}(\rho)$] in the
higher-density ``phase".
%{\color{red}[fstarr: did we answer this question? qualitatively, I feel that a
%``narrower'' basin should have a SMALLER entropy.  narrower = less phase
%space, right?  the converse seems true here.  ice has a higher entropy
%than the amorphous system?  what am I missing?]}

The right column of Fig.~\ref{PEL-LDAHDAICE} shows the PEL properties at
$T=280$~K.  At this temperature the system starts in the liquid state
during compression. For the compression rate employed, the system at
this temperature is essentially in equilibrium, as demonstrated by the fact that the PEL
properties overlap during compression and decompression.  
The main
effect of increasing the temperature from $T=160$~K~$<T_{LLCP}$ to
$T=280$~K~$>T_{LLCP}$ is to lose the  
signature of a distinct 
megabasin associated with the HDL-like liquid.  However, all the PEL features of the HDA-ice transition observed in the left panels of Fig.~\ref{PEL-LDAHDAICE} also occur in the liquid-ice transition shown in the right panels.
%(evidence of such a HDL/HDA megabasin requieres lower temperatures, see Sec.~\ref{compDecomp}).  
%Specifically, Fig.~\ref{PEL-LDAHDAICE}(d) does not show a distinct minimum in $E_{IS}(\rho)$ for liquid states at  $1.0 < \rho <1.45$~g/cm$^3$, that would indicate the transition from  the LDL to the HDL megabasin.  
%A region of negative curvature in $E_{IS}(\rho)$ occurs at $\rho \approx 1-1.45$~g/cm$^3$ and separates the liquid and ice megabasins.
%In addition, there is no van der Waals-like loop in $P_{IS}(\rho)$ that could be associated to an LDL-HDL transformation because $T=280$~K~$>T_{LLCP}$. 
%Instead, it is only during the liquid-ice transition that we find a van der Waals-like loop in $P_{IS}(\rho)$.

In summary, these results suggest that during a true first-order phase
transition, $P_{IS}(\rho)$ shows a van der Waals-like loop, and 
$E_{IS}(\rho)$ exhibits a region of negative curvature.
Interestingly, in the case of the liquid-to-ice transformation, the system also crosses
a potential energy barrier in the PEL that separates two distinct (liquid and ice) 
megabasins.
%This is the same behavior that accompanies the
%LDA-HDA transformation, providing support for the view that 
%the LDA-HDA transformation is the out-of-equilibrium echo of the LDL-HDL first-order phase transition.
The similarities in the PEL properties sampled during the liquid-to-ice and LDA-HDA
 transformations provide support for the view that 
the LDA-HDA transformation is an out-of-equilibrium first-order phase transition;
the origins of the LDA-HDA and LDL-HDL transformations being the system moving
between the same two (LDL/LDA and HDL/HDA) megabasins of the PEL.

\section{LDA, HDA, and the Equilibrium Liquid in the PEL}
\label{liqglassPEL}

We now consider how the LDA and HDA configurations compare with the
configurations sampled by LDL and HDL in equilibrium. This question has
deep implications for our understanding of glass and liquid polymorphism.
For example, if the PEL basins sampled by the LDA (HDA) glass are the same as the ones 
sampled by the equilibrium LDL (HDL) liquid at some fictive temperature $T_f$, then it
becomes possible to provide a thermodynamic modeling of the LDA-HDA
transformation in terms of the LDL-HDL first-order phase transition. 
%One may wonder how the LDA and HDA configurations compare with the
%configurations sampled by the liquid in equilibrium. 
%Specifically, only when LDA/HDA can be associated with a frozen LDL/HDL,
%{\color{red} e.g. when the PEL basins sampled by the glass are the same
%  as the ones sampled by the equilibrium liquid (at a "fictive" $T$),
%  This implies that the relations ${\cal S}_{IS}$ vs $E_{IS}$ and
%  $P_{IS}$ vs $E_{IS}$ must be the same both in the equilibrium liquid
%  and in the glass (at the same $E_{IS}$ ) When this is the case} it
%becomes possible to provide a thermodynamic modelling of the LDA-HDA
%transformation in terms of the LDL-HDL first-order phase transition.  
We addressed this question in Ref.~\cite{yoSPCE-LDAHDA} for the case of
SPC/E water and found that, surprisingly, during the LDA-HDA
transformation the system explores regions of the PEL never sampled by
the equilibrium liquid.  In this section, we show that this
conclusion also holds for the case of ST2 water.  

\begin{figure}[ht]      % Fig4
\centerline  {  \includegraphics[width=9.0cm]{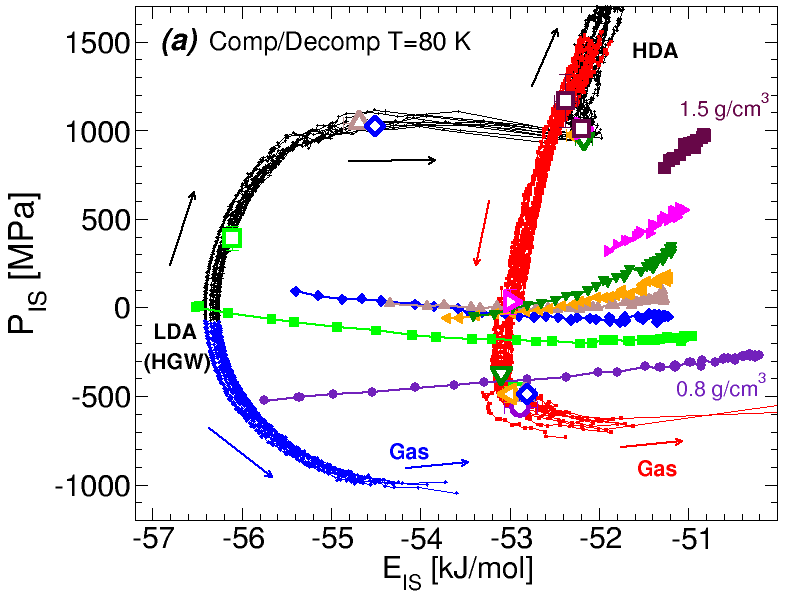}   }
\centerline  {  \includegraphics[width=9.0cm]{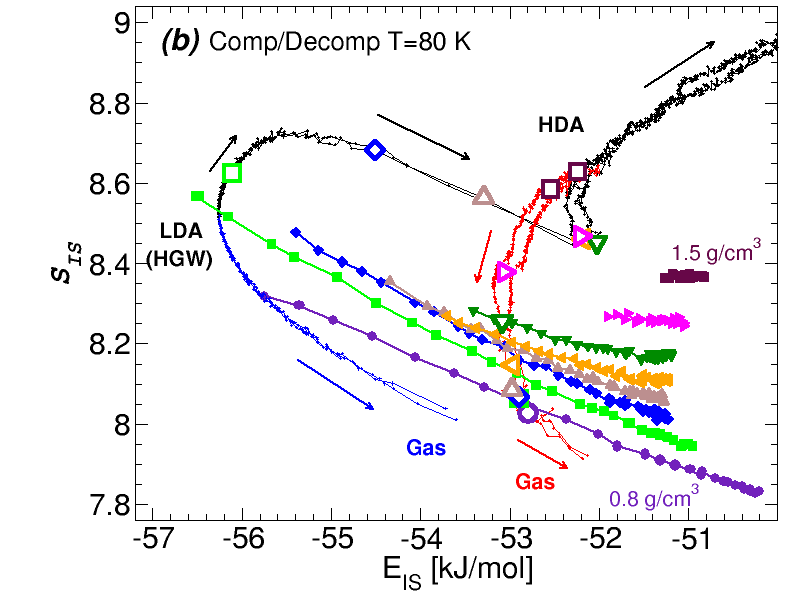}   }
\caption{(a) $P_{IS}$ and (b) ${\cal S}_{IS}$ as functions of $E_{IS}$ during the 
compression/decompression-induced LDA-HDA transformations at $T=80$~K.  Black, red, 
and blue lines correspond, respectively, to the LDA(HGW)-to-HDA, 
HDA-to-LDA-to-gas, and HGW-to-gas transformations.  Filled  symbols are the IS properties of the 
equilibrium liquid at densities $0.8$ (indigo circles),
$0.9$ (green squares), $1.0$ (blue diamonds), $1.1$ 
(brown up-triangles), $1.2$ (orange left-triangles),
 $1,3$ (green down-triangles), $1.4$ (magenta right-triangles) and $1.5$~g/cm$^3$
 (brown squares).
The same (average) densities are indicated along the compression (black lines) 
and decompression (red lines) runs by the corresponding empty symbol.
The relative location of the empty symbols with respect to the corresponding filled 
symbols indicates the different regions of the PEL explored 
by the glasses during the LDA-HDA transformations 
and the equilibrium liquid. Data for the equilibrium liquid is obtained from 
independent simulations of $N=1728$ molecules. 
Results for $P_{IS}$ and $E_{IS}$ in the glass state 
are from $10$ independent compression/decompression runs; $2$ independent runs are used for the calculation of ${\cal S}_{IS}$.
}
\label{PisEisSis-T80}
\end{figure}

\begin{figure}[ht]      % Fig5
\centerline  {
\includegraphics[width=8.0cm]{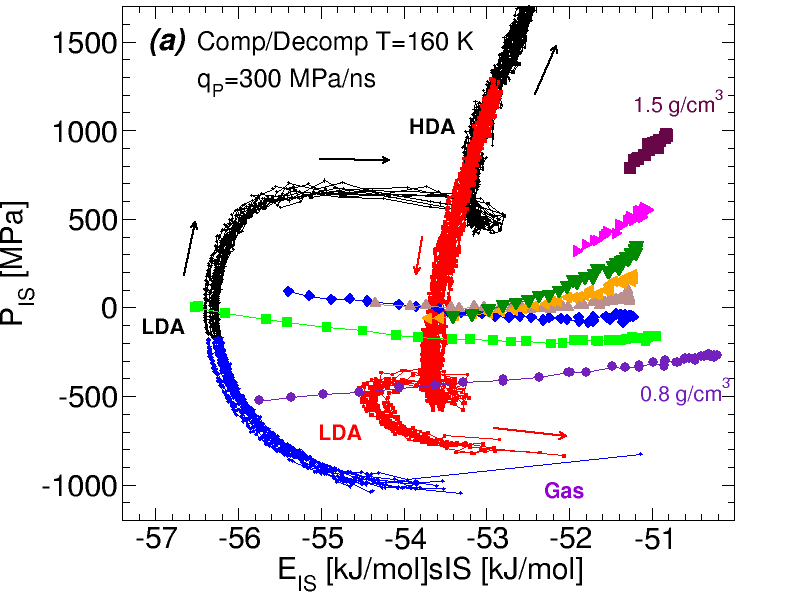}   
\includegraphics[width=8.0cm]{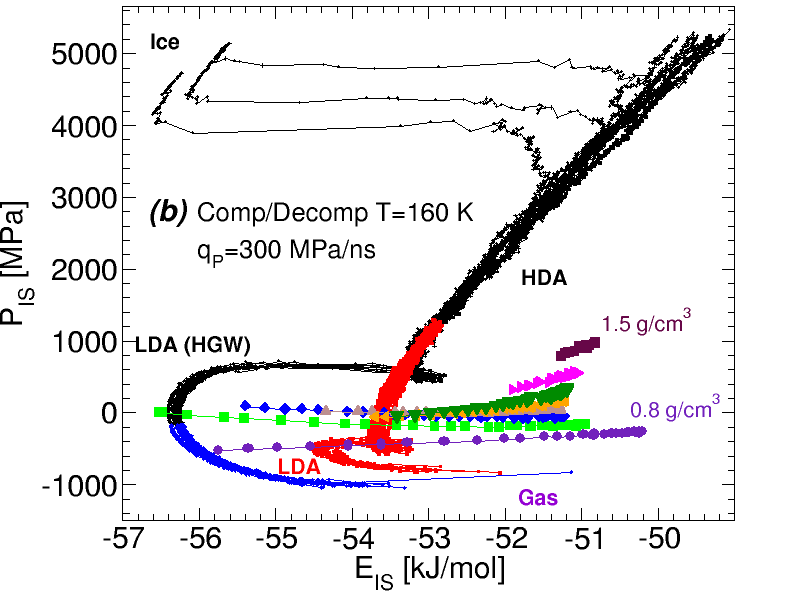}
}
\centerline  {  
\includegraphics[width=8.0cm]{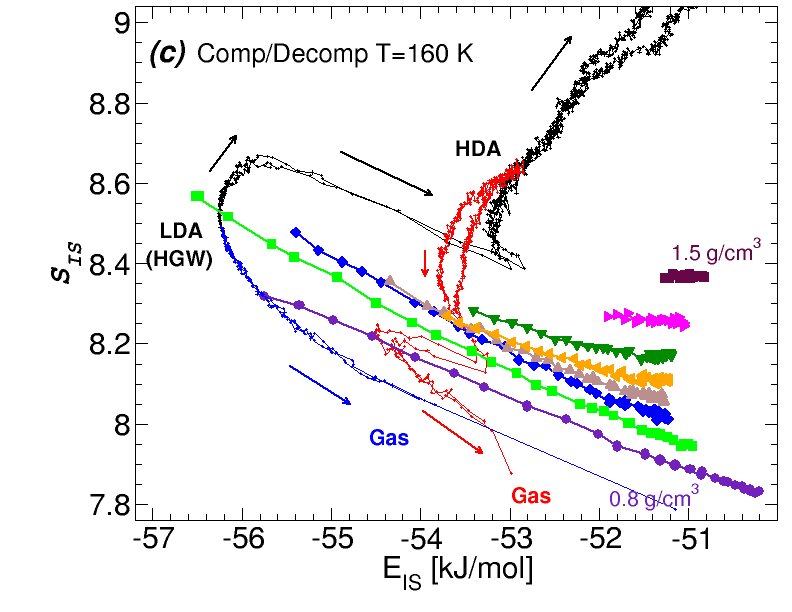}   
\includegraphics[width=8.0cm]{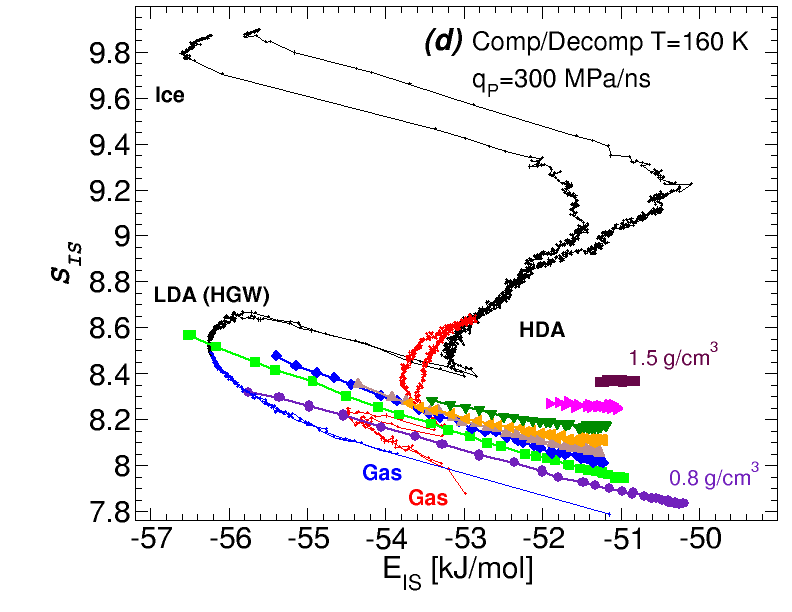}
}
\caption{Same as Fig.~\ref{PisEisSis-T80} for $T=160$~K at low pressures (left panels),
 where crystallization is absent, and at high pressures (right panels),
 where crystallization occurs
in $3$ of the $10$ independent runs. Crystallization is signaled by the 
sharp decrease in the IS energy  in (b), at $P_{IS}>3000$~MPa, and in
 (d), at ${\cal S}_{IS} > 8.8$.
}
\label{PisEisSis-T160}
\end{figure}

To compare the regions of the PEL associated with LDA, HDA, and the
equilibrium liquid, in Figs.~\ref{PisEisSis-T80} and \ref{PisEisSis-T160} we present these states in the $P_{IS}$-$E_{IS}$ and ${\cal S}_{IS}$-$E_{IS}$ planes at 
$T=80$ and $160$~K.
See Fig.~3 of Ref.~\cite{yoSPCE-LDAHDA} for a comparison with the case of
SPC/E water.  The starting LDA form ({\it i.e.} HGW) is close to the
low-energy end of the $\rho=0.9$~g/cm$^3$ isochores for the liquid in the
$P_{IS}$-$E_{IS}$ and ${\cal S}_{IS}$-$E_{IS}$ planes. This is expected
since the density of LDA in this temperature range is
approximately $0.87$~g/cm$^3$.  However, as soon as the compression starts,
the glass deviates abruptly from the liquid isochores, exploring regions
of the PEL that are not accessed by the liquid, similar to the behavior in SPC/E
water~\cite{yoSPCE-LDAHDA}.  For example, at $T=80$~K
the LDA form at $\rho=1.0$~g/cm$^3$ is
characterized by 
$P_{IS}=1000$~MPa, $E_{IS}=-54.5$~kJ/mol, and ${\cal S}_{IS}=8.7$ 
(see empty diamond on the black lines in Fig.~\ref{PisEisSis-T80}),
while
the corresponding values for the equilibrium liquid at the same density
and $E_{IS}$  are $P_{IS}=100$~MPa and
${\cal S}_{IS}<8.35$ (see blue solid diamonds).  
Interestingly, the LDA-HDA transformation occurs
in a region of the PEL that is far from the accessible configurations
explored by the liquid at all densities.

During decompression of HDA (red lines in Figs.~\ref{PisEisSis-T80} and \ref{PisEisSis-T160}), the system returns to regions
of the PEL with properties similar to those of the liquid. However, at a given density, the
LDA form exhibits values of $P_{IS}$, $E_{IS}$, and ${\cal S}_{IS}$ that
correspond to the equilibrium liquid at a different density. For example,
Fig.~\ref{PisEisSis-T80}(a) and \ref{PisEisSis-T80}(b) show that the
decompressed glass at $\rho=1.0$~g/cm$^3$ has $P_{IS} \approx -600$~MPa, 
$E_{IS} \approx -53$~kJ/mol, and
${\cal S}_{IS} \approx 8.05$ (see empty diamond on the red
lines), 
while the corresponding values for
the equilibrium liquid at the same density and $E_{IS}$  are approximately 
$P_{IS}=0$~MPa and ${\cal S}_{IS} =8.2$ (blue solid
diamonds).

The results shown in Fig.~\ref{PisEisSis-T160}(a) for $T=160$~K are
similar.  The main difference between the cases $T=80$~ and $160$~K
occur during the decompression paths.  Specifically,
Fig.~\ref{PisEisSis-T80}(a) and (b) show decompression paths (red lines)
where both $P_{IS}$ and ${\cal S}_{IS}$ decrease monotonically with
decreasing $E_{IS}$. Instead, Fig.~\ref{PisEisSis-T160}(a) and (c) show a
sharp kink at low $P_{IS}$ and ${\cal S}_{IS}$ corresponding to the
HDA-LDA transformation. Yet, at both temperatures the LDA form at a
given density is in a different region of the PEL than the equilibrium liquid at the same density.

For comparison, we include in Figs.~\ref{PisEisSis-T160}(a)-(d) the
location in the $P_{IS}$-$E_{IS}$ and ${\cal S}_{IS}$-$E_{IS}$
planes of the high-pressure ice that forms during compression at
$T=160$~K.  Clearly, the PEL region explored by the system in the ice
state is very different from the regions associated with LDA, HDA, or the
equilibrium liquid.

\section{Discussion}
\label{summary}

In summary, we find that the PEL properties in our simulations of amorphous solid water
 during compression and decompression support the view that LDA and HDA correspond to two
 distinct megabasins in the PEL.  We also show that the PEL behavior we observe in 
the LDA-HDA transformation is qualitatively the same as the PEL signatures observed 
during unambiguous first-order phase transitions in the same system, specifically the 
%HDA-ice and
 liquid-ice transitions.  This finding is consistent with the interpretation 
of the LDA-HDA transformation as a sub-glass-transition manifestation of the equilibrium 
LLPT  that has been demonstrated to occur in ST2 water.  
Comparison of our ST2 results with those obtained using SPC/E water demonstrate 
that when the LLPT is inaccessible, the phase-transition-like characteristics of the LDA-HDA transformation are correspondingly weakened, or lost entirely.
Given the closer similarity between the behavior of amorphous solid ST2 water and real amorphous solid water, our results therefore support the possibility that a LLPT occurs in deeply supercooled liquid water.

%%%%%%%%%%  LDL-HDL along different IS  than LDA-HDA....same megabasins change
An interesting result of this work is that, in ST2 water, 
 the regions of the PEL sampled by the 
liquid (LDL and HDL) and the glass (LDA and HDA) differ. This is in 
agreement with previous simulations of SPC/E water.  It follows that,
at least for the compression/decompression rates explored here, 
the concept of fictive temperature cannot be used to associate LDA or HDA with 
``frozen" equilibrium configurations of LDL or HDL.  As a consequence, it seems 
unlikely that a simple thermodynamic modeling of the LDA-HDA transformation in terms
of the LDL-HDL first-order phase transition is possible.

\begin{figure}[ht]      % Fig6
\centerline  {
\includegraphics[width=10.0cm]{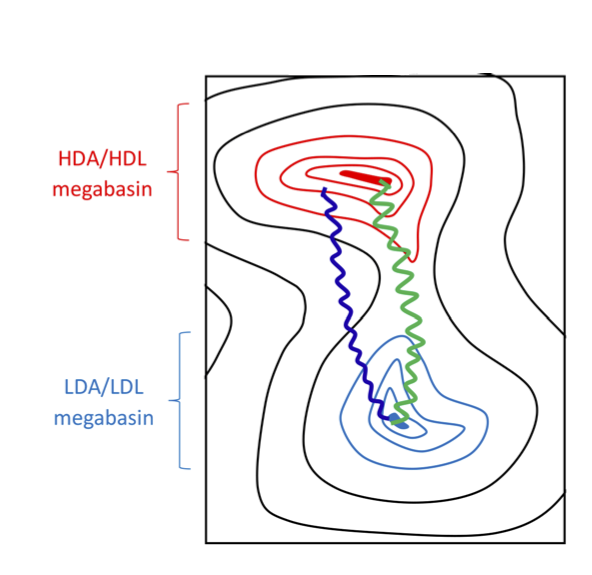}
}
\caption{Schematic of the PEL for ST2 water showing the LDA/LDL (blue) and
HDA/HDL (red) megabasins. The LDA (HDA) configurations are represented by the
solid blue (red) domains and correspond to the deepest configurations of the LDA/LDL
(HDA/HDL) megabasin.  The green and blue paths represent two trajectories followed
by the system during the LDA-HDA transformation at slow and fast compression rate, respectively.
IS are distributed all over the PEL but are omitted for clarity.
}
\label{figYY}
\end{figure}

%{\color{blue}
However, we also note that the fact that LDL/HDL and LDA/HDA sample different regions 
of the PEL is not at odds with the view that there are two distinct megabasins in the PEL, 
one associated with both LDL and LDA, and another megabasin corresponding to both 
HDL and HDA.  A schematic and highly idealized PEL for real water, based on our results 
for ST2 water, is shown in Fig.~\ref{figYY}. 
 Only the LDL/LDA and HDL/HDA megabasins are illustrated; for clarity we have omitted 
the individual IS basins that are distributed all over the PEL.  In Fig.~\ref{figYY},
 the region of the PEL explored by a given realization of a glass (LDA or HDA) should 
be small compared to the region accessible to the liquid.  The slower a liquid is 
cooled toward the glass state, the deeper the system gets in the PEL in the final 
glassy state.  Accordingly, the well-annealed LDA and HDA samples prepared in experiments
 are expected to lie quite close to the the minima of their respective megabasins.  
The region of the PEL explored by the equilibrium liquid will include many higher 
energy configurations.  The path followed in the PEL when compressing at experimental rates 
from LDA to HDA might look something like the green path in Fig.~\ref{figYY}.  
 However, if a HDA sample is prepared by compression of LDA at a very fast rate compared
 to experiments, as is done in our simulations, the glass can be driven into regions of
 the PEL quite different, and at higher energy, than those explored by either the liquid,
 or the well-relaxed HDA obtained in experiments; a possible realization of this scenario is shown
 by the blue path in Fig.~\ref{figYY}.   From this point of view, we should not be surprised
 that different regions of the PEL are sampled by the glass during compression, relative to
 the liquid.  A similar argument has been used to explain the differences between HDA samples
 prepared under different procedures~\cite{mishimaMegabasin}.

One of the limitations of simulation studies of glass polymorphism is the fast 
compression/decompression rates employed, relative to experiments.  In particular, we have shown that the starting LDA form for our compression/decompression cycle (HGW) 
is not identical to the LDA form recovered after decompression of HDA. 
However, as shown in the Appendix, the difference between these two LDA forms decreases upon reducing the compression/decompression rate. Hence the present results 
are consistent with HGW and the recovered LDA form belonging to the same LDL/LDA megabasin.   
Our work supports the interpretation that the LDL/LDA and HDL/HDA megabasins each host a family of glasses and liquids, all with similar properties. 
In a following work, we will expand on the reversibility of the pressure-induced LDA-HDA transformation in simulations using the PEL approach.

\section*{Acknowledgments}

This project was supported, in part, by a grant of computer time from
the City University of New York High Performance Computing Center under
NSF Grants CNS-0855217, CNS-0958379 and ACI-1126113. PHP thanks NSERC
and ACEnet.  We thank Wesleyan University for computational resources.

%%%%%%%%%%%%%%%%%%%%%%%%%%%%%%%%%%%%%%%%%%%%%

\bigskip
\section*{Appendix: Influence of the Compression/Decompression Rate on the PEL Properties Sampled 
during the LDA-HDA Transformations}
\label{sectionAppendix}

In this section we study the effects of reducing the compression/decompression 
rate on the behavior of $E_{IS}(\rho)$, $P_{IS}(\rho)$, and ${\cal S}_{IS}(\rho)$ during 
the LDA-HDA transformations.  We find that the qualitative PEL behavior is not altered by reducing $q_P$ from $300$ to $30$~MPa/ns.
 Interestingly, while the compression-induced LDA-to-HDA transformation is 
 clear at both rates studied, 
the decompression-induced HDA-to-LDA transformation becomes much more
 evident when $q_P$ is reduced. 
In addition, the recovered LDA becomes closer to the starting LDA (HGW) as $q_P$ decreases. 

\begin{figure}[ht]      % Fig7
\centerline  {
\includegraphics[width=7.0cm]{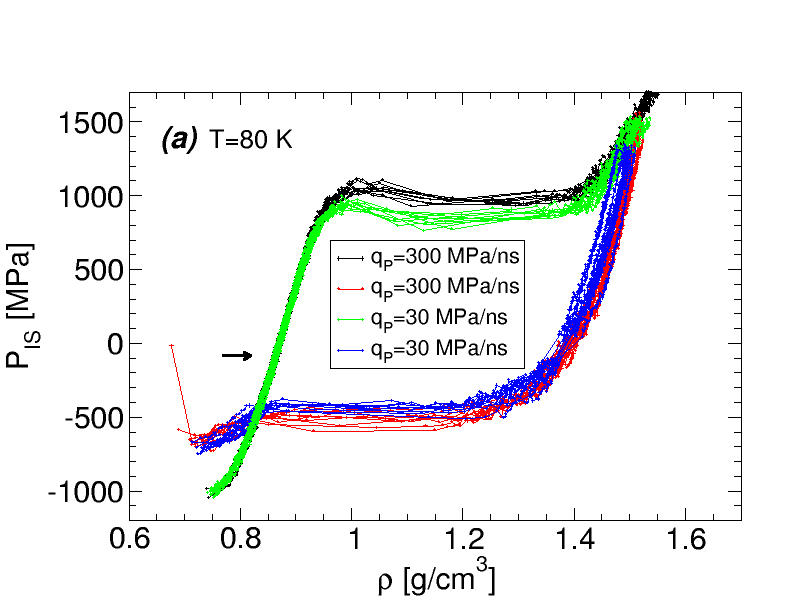} 
\includegraphics[width=7.0cm]{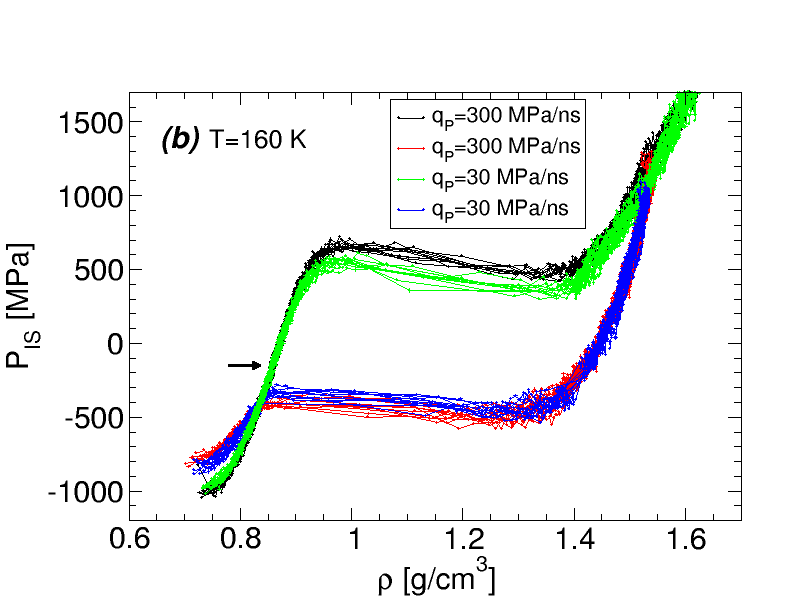}
}
\centerline  {
\includegraphics[width=7.0cm]{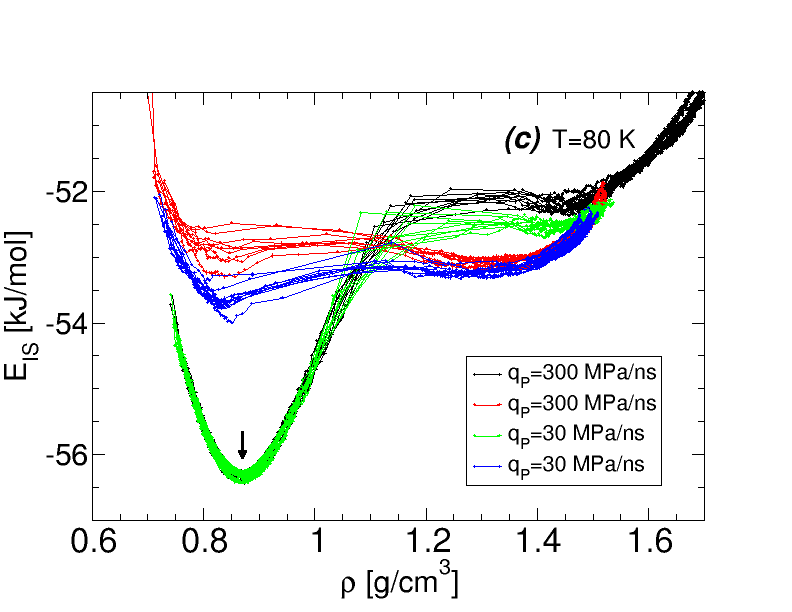}
\includegraphics[width=7.0cm]{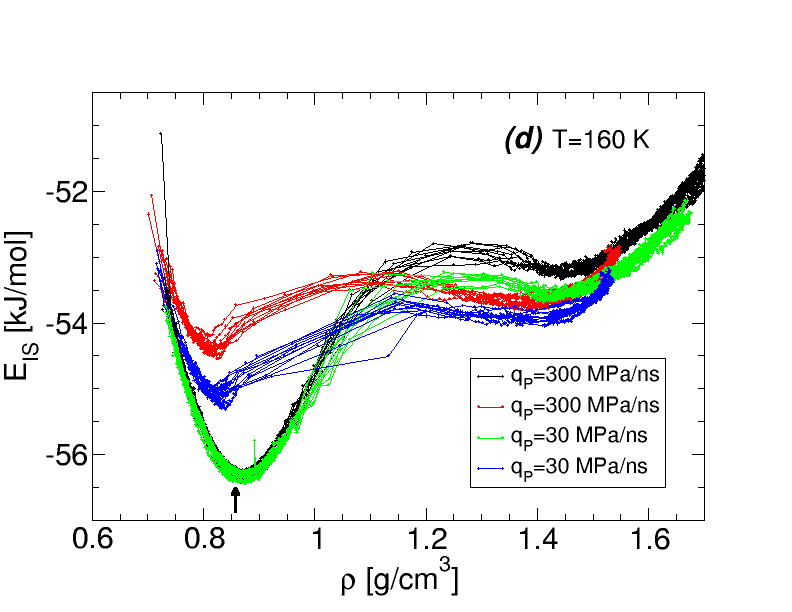}
}
\centerline  {
\includegraphics[width=7.0cm]{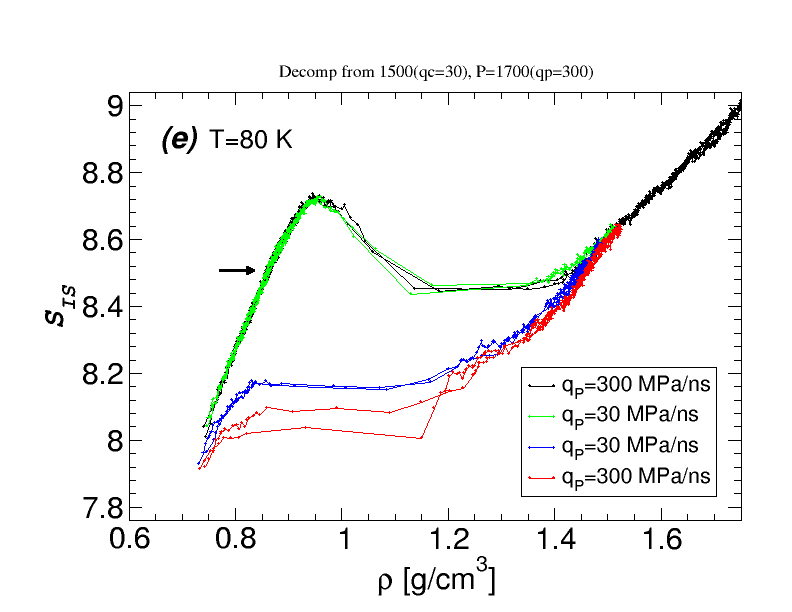}
\includegraphics[width=7.0cm]{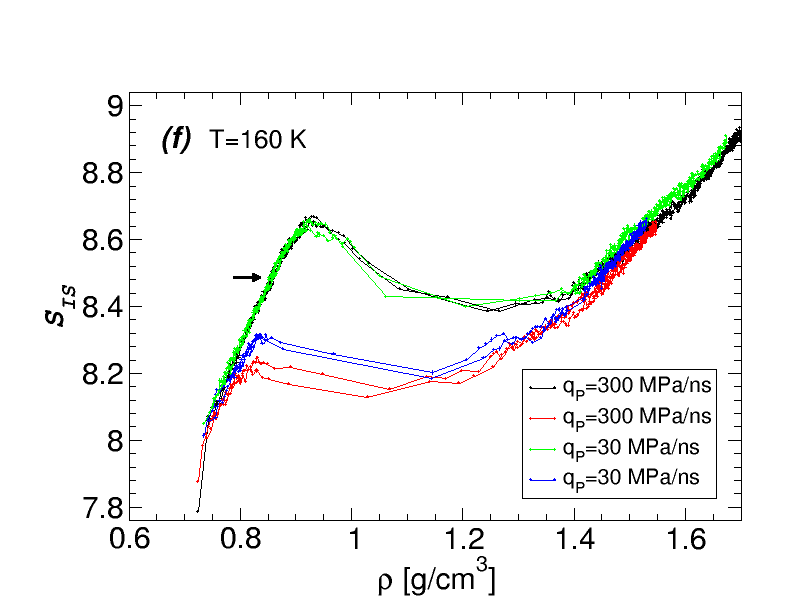}
}
\caption{Effects of the compression/decompression rate ($q_P$) on 
$P_{IS}(\rho)$, $E_{IS}(\rho)$, and shape function 
${\cal S}_{IS}(\rho)$ during the LDA(HGW)-HDA transformations at $T=80$ (left panels) 
and $T=160$~K (right panels).  The starting LDA(HGW) at $P=0.1$~MPa is indicated by the black arrow. 
Green and black lines represent the (i) compression-induced 
LDA-to-HDA transformation, for $P>0.1$~MPa, combined with 
 (ii) the decompression of HGW for the cases $q_P=30$ and $300$~MPa/ns, 
respectively. Blue and red lines correspond to the decompression-induced HDA-to-LDA-to-gas 
transformations at $P<0.1$~MPa, for the cases $q_P=30$ and $300$~MPa/ns, respectively.
 Data for $q_P=300$~MPa/ns is taken from Figs.~\ref{PEL-LDAHDA} ($T=80$~K) and \ref{PEL-LDAHDAICE} 
($T=160$~K).  Reducing $q_P$ does not alter the qualitative behavior of 
$P_{IS}(\rho)$, $E_{IS}(\rho)$, ${\cal S}_{IS}(\rho)$ during the LDA-HDA transformations.
}
\label{PisEisSis_rates}
\end{figure}

Fig.~\ref{PisEisSis_rates} shows $E_{IS}(\rho)$, $P_{IS}(\rho)$, and 
${\cal S}_{IS}(\rho)$ during the pressure-induced LDA-to-HDA transformation
 at $T=80$~K and $T=160$~K, for
both $q_P=30$ and $300$~MPa/ns. During compression, reducing $q_P$ has the main effect of reducing 
the values of  $E_{IS}(\rho)$, $P_{IS}(\rho)$, and ${\cal S}_{IS}(\rho)$ during
the LDA-HDA transformation. Yet at both rates one observes the signatures
expected during the transition from one megabasin (LDA) to another (HDA)
upon compression: 
(i)  $P_{IS}(\rho)$ shows a van der  Waals-like loop;
(i) $E_{IS}(\rho)$ exhibits a region of negative curvature;
and (iii) ${\cal S}_{IS}(\rho)$ decreases sharply.  

We observe a larger effect during the decompression path when $q_P$ is reduced. 
As discussed in Sec.~\ref{LDAHDA},
the HDA-LDA transformation is rather smooth at $T=80$~K.  The van der Waals-like loop in $P_{IS}(\rho)$, and the negative curvature in $E_{IS}(\rho)$, are weaker effects during decompression, relative to the compression process.
However, upon reducing the decompression rate 
to $q_P=30$~MPa/ns, we observe that these features become significantly more prominent.
This observation is consistent with our conclusion that during decompression, the system 
transitions from the HDA megabasin back to the starting LDA megabasin of the compression cycle.
In particular, Figs.~\ref{PisEisSis_rates}(c) and \ref{PisEisSis_rates}(d)  show
that the value of  $E_{IS}(\rho)$ for the recovered LDA at $\rho=0.87$~MPa decreases 
with decreasing $q_P$ and becomes closer to the IS energy of the starting LDA (HGW). 
That is, the recovered LDA 
is deeper in the LDA megabasin with decreasing compression/decompression rate,
 as one would expect.  This supports the view that the LDA-HDA transformation in ST2 water is indeed reversible~\cite{chiu1}.

\begin{figure}[ht]      % Fig8
\centerline  {
\includegraphics[width=8.0cm]{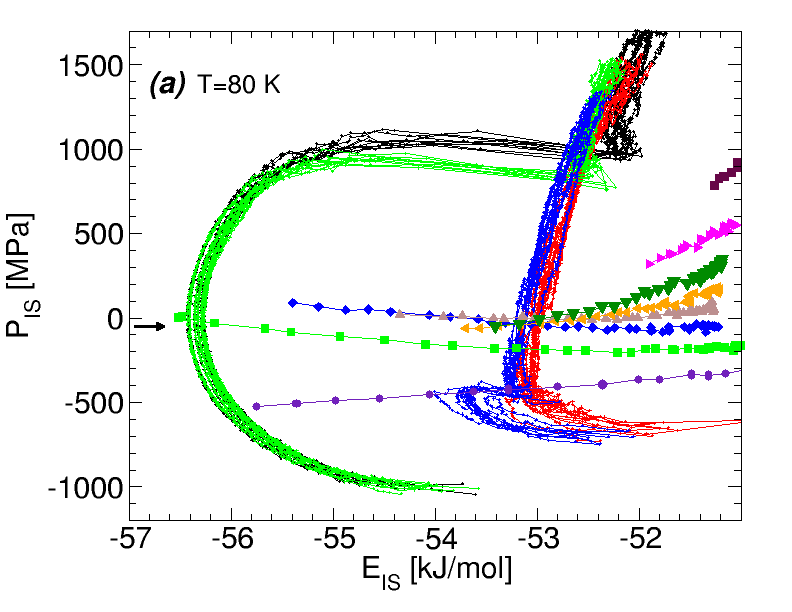}
\includegraphics[width=8.0cm]{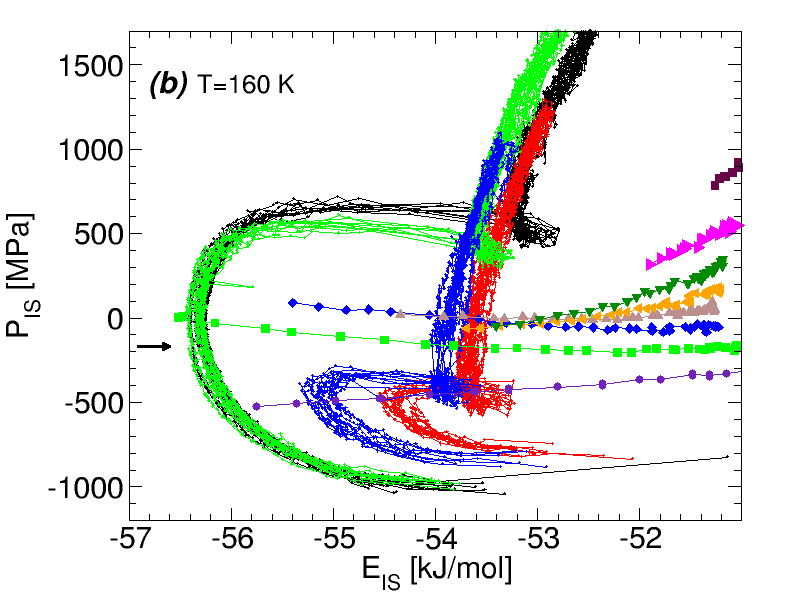}
}
\centerline  {
\includegraphics[width=8.0cm]{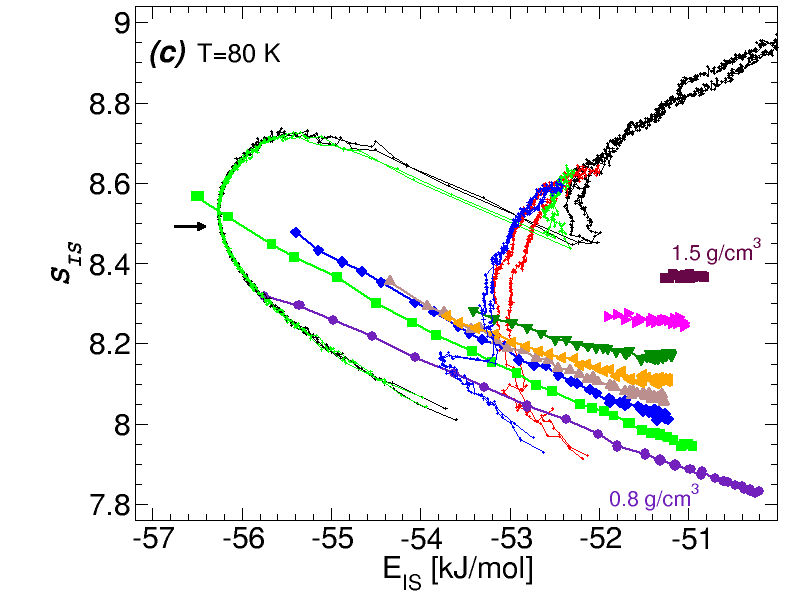}
\includegraphics[width=8.0cm]{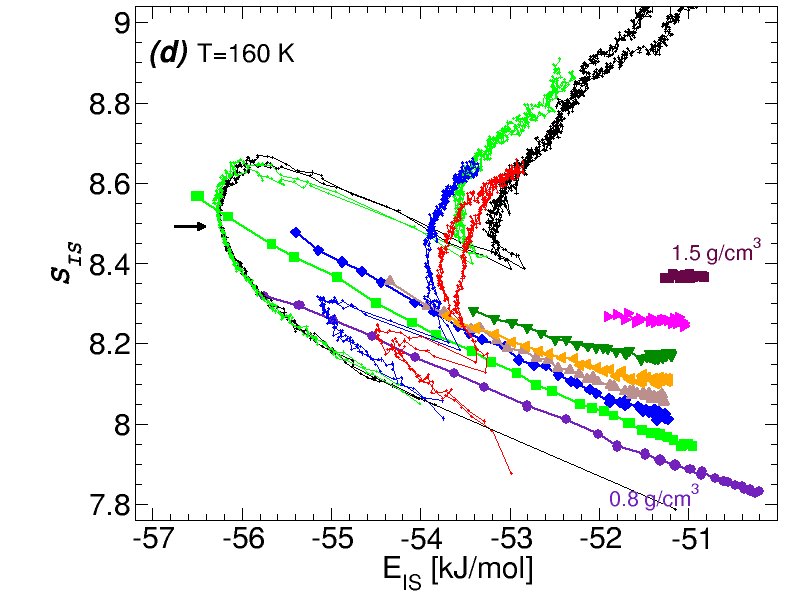}
}
\caption{Effects of the compression/decompression rate ($q_P$)
 on $P_{IS}$ and ${\cal S}_{IS}$ during the LDA-HDA transformation at $T=80$~K 
(left panels) and $T=160$~K (right panels). Data is taken from Fig.~\ref{PisEisSis_rates}.
Crystallization at $T=160$~K occurs at high pressures and is not shown. 
The starting LDA(HGW) at $P=0.1$~MPa is indicated by the black arrow.
Green and black lines represent the (i) compression-induced
LDA-to-HDA transformation, for $P>0.1$~MPa, combined with
the (ii) decompression of HGW at $P<0.1$~MPa, for the cases $q_P=30$ and $300$~MPa/ns,
respectively. Blue and red lines correspond to the decompression-induced HDA-to-LDA-to-gas
transformations for the cases $q_P=30$ and $300$~MPa/ns, respectively.
At both rates studied, the IS sampled by the glasses during the LDA-HDA transformations 
are not accessible to the equilibrium liquid.
}
\label{PisEisSisEis_rates}
\end{figure}

Fig.~\ref{PisEisSisEis_rates} shows $P_{IS}(E_{IS})$ and ${\cal S}_{IS}(E_{IS})$ 
at $T=80$ and $160$~K for $q_P=30$ and $300$~MPa/ns. At both rates studied, 
these properties exhibit the same qualitative behavior during 
compression (black and green lines) and decompression (red and blue lines). 
The main point of Fig.~\ref{PisEisSisEis_rates} is that the system samples IS never visited by 
the equilibrium liquid even if the compression/decompression rate is reduced to $q_P=30$~MPa/ns.

%%%%%%%%%%%%%%%%%%%%%%%%%%%%%%%%%%%%%

%\end{document}

%\newpage

%\vspace{1cm}
%\begin{table}[ht]
%  \caption{    }
%\label{table1}
%\begin{center}
%\begin{tabular}{|c|c|c|}
%\hline
%Compression $T$ [K] & $n$ & $P_{\rm  HDA \rightarrow VII}$ [MPa]\\
%\hline
%140 & 0 & -\\
%\hline
%160 & 3 & 4780\\
%\hline
%180 & 3 & 3652\\
%\hline
%200 & 8 & 2188\\
%\hline
%210 & 9 & 1828\\
%\hline
%220 & 6 & 1456\\
%\hline
%230 & 10 & 869\\
%\hline
%240 & 10 & 812\\
%\hline
%250 & 10 & 765\\
%\hline
%260 & 10 & 819\\
%\hline
%270 & 10 & 968\\
%\hline
%280 & 10 & 1041\\
%\hline
%\end{tabular}
%\end{center}
%\end{table}

%\vspace{1cm}

%\newpage

\newpage

\newpage

\newpage

\newpage

\newpage

\newpage

\newpage

\end{document}